\documentclass[useAMS,usenatbib,usegraphicx]{mn2e}

\newcommand{\kms}{km s$^{-1}$}

\newcommand{\lam}{$\lambda$}
\newcommand{\civ}{\mbox{C\,{\sc iv}}}

\newcommand{\cii}{\mbox{C\,{\sc ii}}}
\newcommand{\siiv}{\mbox{Si\,{\sc iv}}}

\newcommand{\al}{\mbox{Al\,{\sc iii}}}
\newcommand{\oi}{\mbox{O\,{\sc i}}}

\title[Variability in quasar BAL outflows II]
{Variability in quasar broad absorption line outflows II. Multi-epoch monitoring of \siiv\ and \civ\ BAL variability}
\author[D. M. Capellupo et al.]{D. M. Capellupo$^{1}$
\thanks{E-mail:dancaps@astro.ufl.edu (DMC)}, F. Hamann$^{1}$, J. C. Shields$^{2}$, P. Rodr\'iguez Hidalgo$^{3}$,\newauthor and T.A. Barlow$^{4}$\\
$^{1}$Department of Astronomy, University of Florida, Gainesville, FL 32611-2055\\
$^{2}$Department of Physics \& Astronomy, Ohio University, Athens, OH 45701\\
$^{3}$Department of Astronomy and Astrophysics, Pennsylvania State University, University Park, PA 16802\\
$^{4}$Infrared Processing and Analysis Center, California Institute of Technology, Pasadena, CA 91125}
\begin{document}


\pagerange{\pageref{firstpage}--\pageref{lastpage}} \pubyear{2002}

\maketitle

\label{firstpage}

\begin{abstract}
Broad absorption lines (BALs) in quasar spectra indicate high-velocity outflows that may be present in all quasars and could be an important contributor to feedback to their host galaxies. Variability studies of BALs help illuminate the structure, evolution, and basic physical properties of the outflows. Here we present further results from an ongoing BAL monitoring campaign of a sample of 24 luminous quasars at redshifts 1.2 $< z <$ 2.9. We directly compare the variabilities in the \civ\ $\lambda$1549 and \siiv\ $\lambda$1400 absorption to try to ascertain the cause(s) of the variability. We find that \siiv\ BALs are more likely to vary than \civ\ BALs. When looking at flow speeds $>$$-$20 000 \kms, 47 per cent of quasars exhibited \siiv\ variability while 31 per cent exhibited \civ\ variability. Furthermore, $\sim$50 per cent of the variable \siiv\ regions did not have corresponding \civ\ variability at the same velocities, while nearly all occurrences of \civ\ variability had corresponding changes in \siiv. We do not find any correlation between the absolute change in strength in \civ\ and in \siiv, but the fractional change in strength tends to be greater in \siiv\ than in \civ. When both \civ\ and \siiv\ varied, those changes always occurred in the same sense (either getting weaker or stronger). We also include our full data set so far in this paper, which includes up to 10 epochs of data per quasar. The multi-epoch data show that the BAL changes were not generally monotonic across the full $\sim$5 to $\sim$8 yr time span of our observations, suggesting that the characteristic time-scale for significant line variations, and (perhaps) for structural changes in the outflows, is less than a few years. Coordinated variabilities between absorption regions at different velocities in individual quasars seems to favor changing ionization of the outflowing gas as the cause of the observed BAL variability. However, variability in limited portions of broad troughs fits naturally in a scenario where movements of individual clouds, or substructures in the flow, across our lines-of-sight cause the absorption to vary. The actual situation may be a complex mixture of changing ionization and cloud movements. Further discussion of the implications of variability, e.g., in terms of the size and location of the outflowing gas, will be presented in a forthcoming paper.
\end{abstract}


\section{Introduction}

High-velocity wind-like outflows are an important part of the quasar system and a potential contributor to feedback to the host galaxy. These outflows may play a crucial role in the accretion process in quasars and the growth of supermassive black holes (SMBHs), by allowing the accreting material to release angular momentum. Furthermore, quasar outflows may provide enough kinetic energy feedback to affect star formation in the quasar host galaxies, to aid in `unveiling' dust-enshrouded quasars, and to help distribute metal-rich gas to the intergalactic medium (e.g., \citealt{DiMatteo05}; \citealt{Moll07}).

Broad absorption lines (BALs) are the most prominent signatures of accretion disk outflows seen in quasar spectra. BALs are defined as absorption troughs with velocity widths $>$2000 \kms\ at depths $>$10\% below the continuum \citep{Weymann91}, and they appear in the spectra of $\sim$10-15\% of quasars (\citealt{Reichard03a}; \citealt{Trump06}; \citealt{Gibson09}). Since BALs are seen in just a fraction of quasar spectra, their presence could represent a phase in the evolution of a quasar and/or particular orientations where the outflow lies between us and the quasar emission sources.

The location and three-dimensional structure of quasar outflows are poorly understood. Sophisticated models predict these outflows as arising from a rotating accretion disk, with acceleration to high speeds by radiative and/or magneto-centrifugal forces (\citealt{Murray95}; \citealt{Proga04}; \citealt{Proga07}; \citealt{Everett05}). Improved observational constraints are necessary to test these models and to estimate mass-loss rates, kinetic energy yields, and the role of quasar outflows in feedback to the surrounding environment.

One way to obtain constraints on quasar outflows is to study the variability in their absorption lines, which can provide information on the structure and dynamics of the outflowing gas. Two possible causes of this observed variability are movement of gas across our line of sight to the quasar and changes in ionization (\citealt{Barlow93}; \citealt{Wise04}; \citealt{Misawa05}; \citealt{Lundgren07}; \citealt{Gibson08}; \citealt{Hamann08}). Variability on shorter time-scales can place constraints on the distance of the absorbing material from the central SMBH. Shorter variability time-scales indicate smaller distances, based on nominally shorter crossing times for moving clouds (\citealt{Hamann08}; \citealt{Capellupo11}) or the higher densities required for shorter recombination times \citep{Hamann97}. Measurements of variability on longer (multi-year) time-scales provide insight into the homogeneity and stability of the outflowing gas. If no variability is detected on long time-scales, then this indicates a smooth flow with a persistent structure. Overall, results of variability studies provide information on the size, kinematics, and internal makeup of sub-structures within the outflows. Furthermore, variability studies can address the evolution of these outflows as absorption lines have been observed to appear and disappear (\citealt{Hamann08}; \citealt{Leighly09}; \citealt{Krongold10}; \citealt{RodriguezH11}), or they can evolve from one type of outflow feature to another (e.g., from a mini-BAL to a BAL or vice versa; \citealt{Gibson10}; Rodr\'iguez Hidalgo et al. in preparation; also this work).

Most of the existing work on BAL variability has focused on variability in \civ\ $\lambda$1549 over two epochs (e.g., \citealt{Barlow93}; \citealt{Lundgren07}; \citealt{Gibson08}). \citealt{Gibson08} detected \civ\ BAL variability in 12 out of 13 BAL quasars (92 per cent) over multi-year time-scales. None of these studies found clear evidence for acceleration in the BALs. \citet{Gibson10} reports on variability on multi-month to multi-year rest-frame time-scales, using 3-4 epochs of data for 9 BALQSOs and found that BALs generally do not vary monotonically over time. Their study also makes comparisons between variability in \siiv\ $\lambda$1400 absorption and variability in \civ, and their results include a correlation between fractional change in EW in \siiv\ and \civ.

This work is the second paper in a series on BAL variability. The first paper, \citeauthor{Capellupo11} (\citeyear{Capellupo11}; hereafter, Paper 1), introduced our ongoing monitoring programme of a sample of 24 BAL quasars. We began with a sample of BAL quasars from \citet{Barlow93}, which includes spectra of the \civ\ absorption region and, in most cases, coverage of the \siiv\ absorption region as well. We have re-observed these quasars to provide a longer time baseline over which to study variability, as well as to obtain multiple epochs of data per object. We currently have up to 10 epochs of data per quasar up to March 2009, covering rest-frame time intervals ($\Delta$t) from 15 days to 8.2 yr\footnote{Throughout this paper, all time intervals are measured in years in the rest frame of the quasar.}.

Paper 1 focused on a subset of the data from this monitoring programme to look for basic trends in the data between variability and other properties of the absorbers, as well as to directly compare short-term and long-term variability within the same sample of quasars. Paper 1 took a novel approach to studying BAL variability by introducing a measurement of BAL strength within portions of a trough, instead of using equivalent width (EW) measurements. Paper 1 discusses variability in just two different time intervals: a short-term interval of 0.35$-$0.75 yr and a long-term interval of 3.8$-$7.7 yr. We found that 39 per cent (7/18) of the quasars varied in the short-term, whereas 65 per cent (15/23) varied in the long-term data. The variability most often occurred in just portions of a BAL trough, which is similar to the findings of \citet{Gibson08}. We found that the incidence of variability was greater at higher velocities and in weaker portions of BAL troughs. Similarly, in \citet{Lundgren07}, the strongest occurrences of BAL variability occurred at velocities $<$$-$12 000 \kms\ and in features with smaller equivalent widths. Overall, the results of Paper 1 are broadly consistent with previous work on BAL variability.

In this paper, we extend the analysis of Paper 1 by looking at variability in \siiv\ and comparing it to the variability results for \civ. Expanding our study to include \siiv\ absorption can help constrain theories on the cause(s) of BAL variability. C and Si have different abundances, if solar abundances are assumed, and they have different ionization properties (e.g. \citealt{Hamann08}, \citeyear{Hamann11}). By examining if \civ\ and \siiv\ have different variability properties, and how they differ, coupled with these differences in abundances and ionization properties, we can gain new insight into the cause(s) of BAL variability.

Our dataset is uniquely suited to this study because we have coverage of the \siiv\ line for nearly our entire sample (22 out of 24 quasars). In addition to the larger sample size, we go beyond existing work by adopting a method of measuring the absorption strength in portions of BALs, instead of EW. Equivalent width measurements apply to an entire feature and are less sensitive to changes in small portions of troughs. Our method of measuring portions of troughs also allows more direct comparisons between the behavior of \civ\ and \siiv\ variability.

We also include the entire dataset so far to look at variability in \civ\ and \siiv\ over multiple epochs. This work contains up to 10 epochs of data per quasar, and including all of these epochs will provide better insight into the characteristics of BAL variability. Increasing the number of epochs provides new information on whether BALs change monotonically over time or whether they can vary and then return to an earlier state. In Paper 1, we reported that typically only portions of BALs varied. Multi-epoch data can tell us whether variability only occurs in those specific velocity intervals or if the velocity range over which variability occurs can change over time. We also highlight several individual interesting cases of variability that can further help us understand BAL outflows. Section 2 below reviews the quasar sample and analysis introduced in Paper 1, Section 3 describes our results, Section 4 summarizes the results so far from Paper 1 and the current work, and Section 5 discusses the results and their implications.

\section[]{Data and Analysis}

\begin{table*}
  \begin{minipage}{110mm}
    \caption{Quasar Data}
    \begin{tabular}{cccccccc}
\hline
Name & $z_{em}$ & BI & Lick & SDSS & MDM & Total & $\Delta$t \\
 & & & 1988-92 & 2000-06 & 2007-09 & & (yrs) \\
\hline
0019+0107	& 2.130 & 2290		& 6 & 0 & 1 & 7 & 0.08-5.79 \\
0043+0048   	& 2.137 & 4330  	& 2 & 2 & 1 & 5 & 0.35-6.13 \\
0119+0310   	& 2.090 & 6070  	& 2 & 0 & 1 & 3 & 0.65-5.57 \\
0146+0142   	& 2.909 & 5780  	& 2 & 0 & 2 & 4 & 0.52-5.15 \\
0226$-$1024 	& 2.256 & 7770  	& 1 & 0 & 1 & 2 & 4.66 \\
0302+1705   	& 2.890 & 0       		& 2 & 0 & 1 & 3 & 0.27-4.42 \\
0842+3431   	& 2.150 & 4430  	& 6 & 1 & 3 & 10 & 0.06-5.87 \\
0846+1540   	& 2.928 & 0       		& 5 & 0 & 3 & 8 & 0.04-4.93 \\
0903+1734   	& 2.771 & 10700	& 2 & 1 & 4 & 7 & 0.04-5.29 \\
0932+5006   	& 1.926 & 7920  	& 4 & 1 & 4 & 9 & 0.05-6.98 \\
0946+3009   	& 1.221 & 5550  	& 2 & 0 & 3 & 5 & 0.11-8.16 \\
0957$-$0535 	& 1.810 & 2670  	& 2 & 0 & 2 & 4 & 0.11-6.21 \\
1011+0906   	& 2.268 & 6100  	& 4 & 0 & 3 & 7 & 0.10-5.94 \\
1232+1325   	& 2.364 & 11000	& 1 & 0 & 2 & 3 & 0.35-5.93 \\
1246$-$0542 	& 2.236 & 4810  	& 2 & 0 & 2 & 4 & 0.40-5.90 \\
1303+3048   	& 1.770 & 1390  	& 1 & 0 & 4 & 5 & 0.05-6.10 \\
1309$-$0536 	& 2.224 & 4690  	& 2 & 0 & 2 & 4 & 0.68-6.19 \\
1331$-$0108 	& 1.876 & 10400 	& 2 & 1 & 2 & 5 & 0.42-5.97 \\
1336+1335   	& 2.445 & 7120   	& 1 & 0 & 3 & 4 & 0.07-5.79 \\
1413+1143   	& 2.558 & 6810  	& 2 & 1 & 2 & 5 & 0.26-5.61 \\
1423+5000   	& 2.252 & 3060  	& 2 & 1 & 2 & 5 & 0.39-5.87 \\
1435+5005   	& 1.587 & 11500 	& 2 & 0 & 2 & 4 & 0.34-7.72 \\
1524+5147   	& 2.883 & 1810  	& 3 & 1 & 3 & 7 & 0.04-5.14 \\
2225$-$0534 	& 1.981 & 7920  	& 3 & 0 & 0 & 3 & 0.27-0.73 \\
\hline
    \end{tabular}
  \end{minipage}
\end{table*}

\subsection{Observations and quasar sample}

In this work, we use the same sample of 24 BAL quasars introduced in Paper 1. This sample is based on the set of BALQSOs studied in \citet{Barlow93}. The sample selection and general characteristics are described in Paper 1. These data were obtained from the Lick Observatory 3-m Shane Telescope, using the Kast spectrograph. Most of the spectra we use from that data set have a resolution of $R\equiv\lambda/\Delta\lambda\approx 1300$ (230 \kms). For epochs where this resolution is not available, we use data taken at $R \approx$ 600 (530 \kms). BALs are defined to have a width of at least 2000 \kms, so either of these resolutions is sufficient to measure the lines and study their variabilities. The wavelength coverage of each spectrum covers at least the \siiv\ through \civ\ emission lines, and most cover at least the Ly$\alpha$ \lam 1216 through \civ\ emission lines.

We have been re-observing 23 of the BALQSOs from \citet{Barlow93} at the MDM Observatory 2.4-m Hiltner telescope, using the CCDS spectrograph with a resolution of $R \approx$ 1200 (250 \kms). The observations used in this work were taken in January and February 2007; January, April, and May 2008; and January and March 2009. We used the same spectrograph setup each time, varying only the wavelength range in order to observe each quasar at roughly the same rest wavelength range, from Ly$\alpha$ through \civ\ emission. One exception is 0946+3009, which has a redshift too low for the \siiv\ emission to appear in our spectra.

We supplement our data with spectra from the SDSS Data Release 6 \citep{Adelman08} for 8 of the quasars in our sample, for which the resolution is $R\approx 2000$ (150 \kms). These spectra cover the observed wavelength range 3800 to 9200 \AA, and we only include spectra that cover at least the \siiv\ through \civ\ emission.

Table 1 summarizes the full dataset presented in this work, including the emission redshift, $z_{em}$,\footnote{The values of $z_{em}$ were obtained from the NASA/IPAC Extragalactic Database (NED), which is operated by the Jet Propulsion Laboratory, California Institute of Technology, under contract with the National Aeronautics and Space Administration.} and the `balnicity index' (BI) for each object (as calculated in Paper 1). Any uncertainty in the redshift will not affect our comparisons between \siiv\ and \civ\ or any of our other results. The balnicity index, defined by \citet{Weymann91}, is a measure of the strength of the BAL absorption and is calculated as an EW in units of velocity. It quantifies blue-shifted \civ\ absorption between $-$25 000 and $-$3 000 \kms\ that reaches at least 10 per cent below the continuum across a region at least 2000 \kms\ in width. The next four columns list the number of observations taken for each object at each observatory and then the total overall number of observations. The final column lists the range in $\Delta$t covered for each quasar.

Two of the quasars in our sample have BI=0, so they are not BAL quasars based on the balnicity index. They both contain broad absorption, but this absorption falls outside the velocity range, $-$25 000 to $-$3 000 \kms, used to define BI. As noted in Paper 1 and discussed further in Section 3 below, including these two objects in our sample does not affect any of our main results.

\subsection{Measuring BALs and their variability}

\begin{figure*}
  \includegraphics[width=152mm]{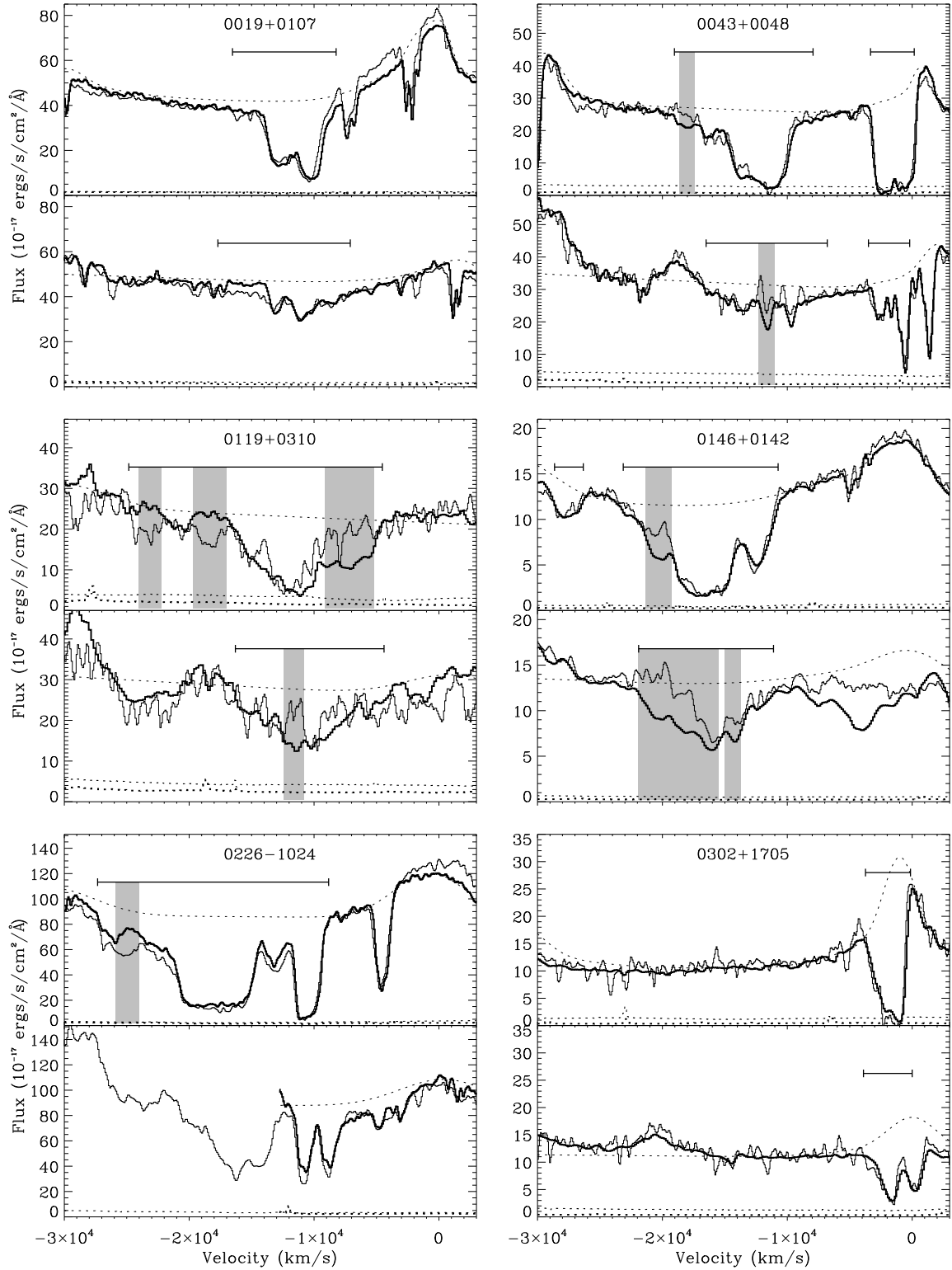}
  \caption{Spectra of all 24 quasars in our sample, showing the long-term comparisons
    ($\Delta$t = 3.8$-$7.7 yr) between a Lick Observatory spectrum (bold curves) and a recent MDM
    spectrum (thin curves). For each quasar, the \civ\ region is displayed in the top panel and
    the corresponding \siiv\ region is shown in the bottom panel. For 2225-0534, we only have
    short-term Lick data (see Table 1). The vertical flux scale applies to the Lick data, and the MDM
    spectrum has been scaled to match the Lick data in the continuum. The dashed curves show our
    pseudo-continuum fits. The horizontal bars indicate intervals of BAL absorption included in this
    study, and the shaded regions indicate intervals of variation within the BALs. We used binomial
    smoothing to improve the presentation of the spectra. The formal 1$\sigma$ errors are shown
    near the bottom of each panel. (The two variability intervals defined for 1524+5147 were labeled
    as one interval in Paper 1.)}
  \label{spectra}
\end{figure*}
\addtocounter{figure}{-1}
\begin{figure*}
  \includegraphics[width=170mm]{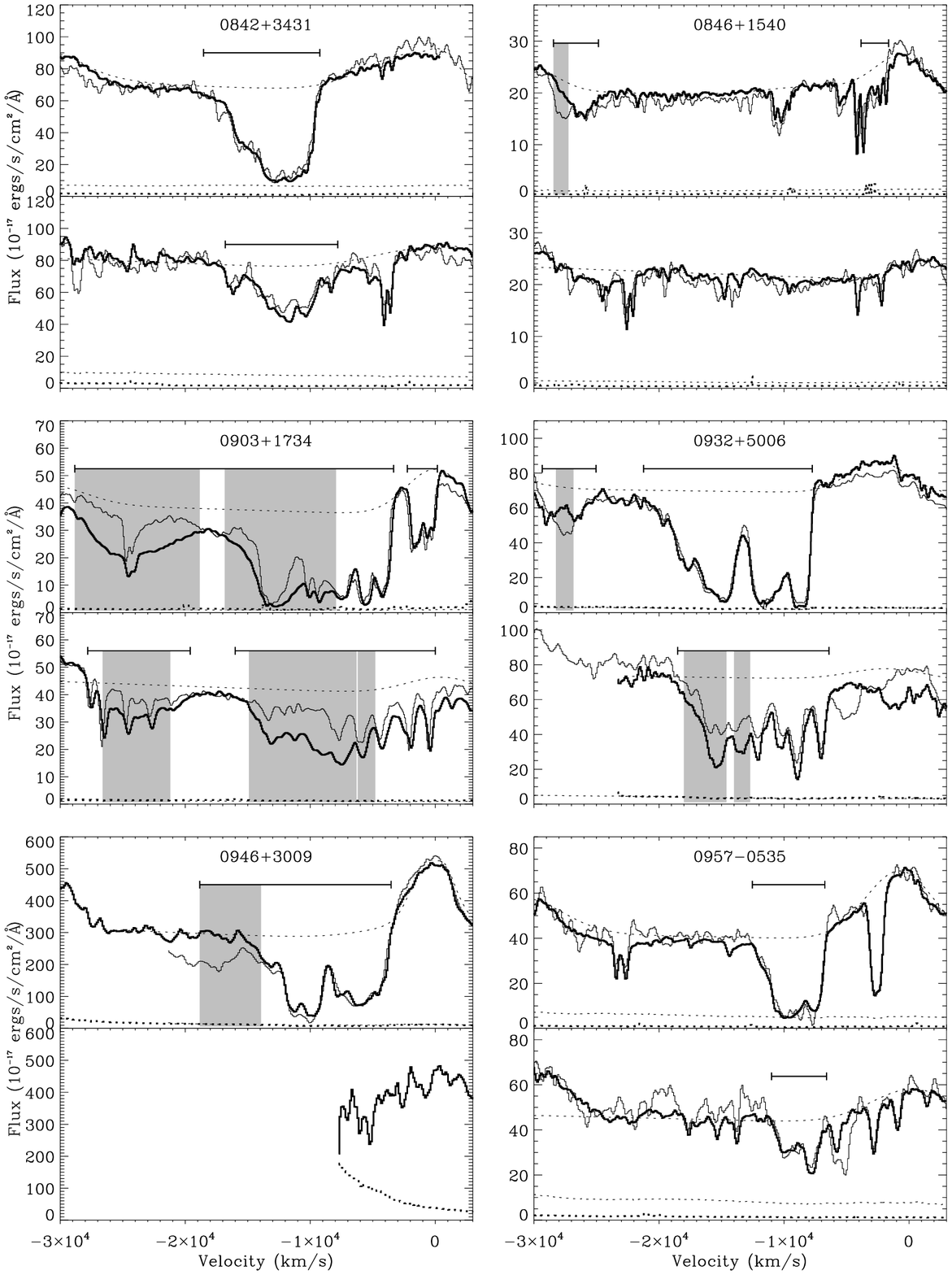}
  \caption{continued...}
\end{figure*}
\addtocounter{figure}{-1}
\begin{figure*}
  \includegraphics[width=170mm]{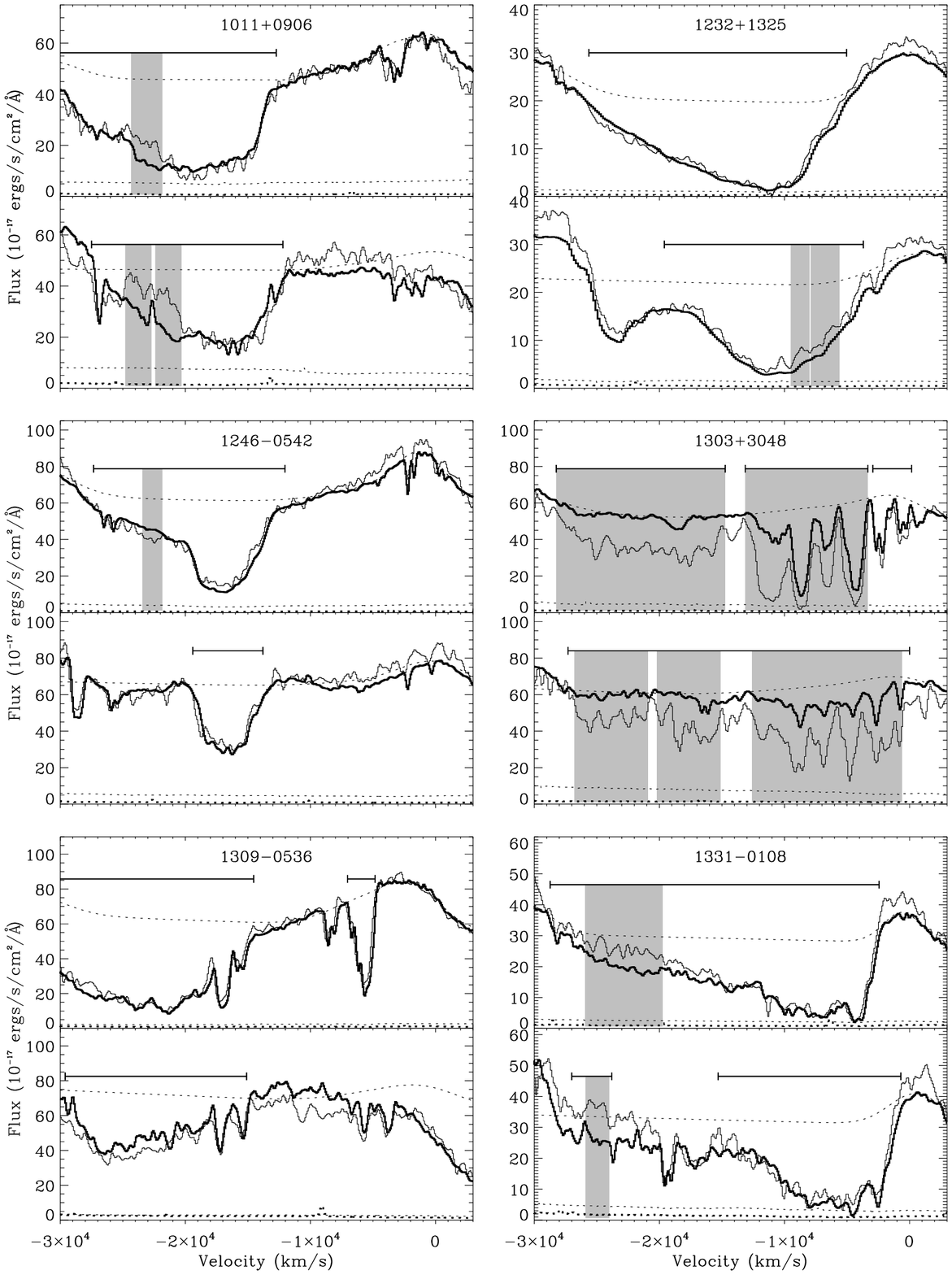}
  \caption{continued...}
\end{figure*}
\addtocounter{figure}{-1}
\begin{figure*}
  \includegraphics[width=170mm]{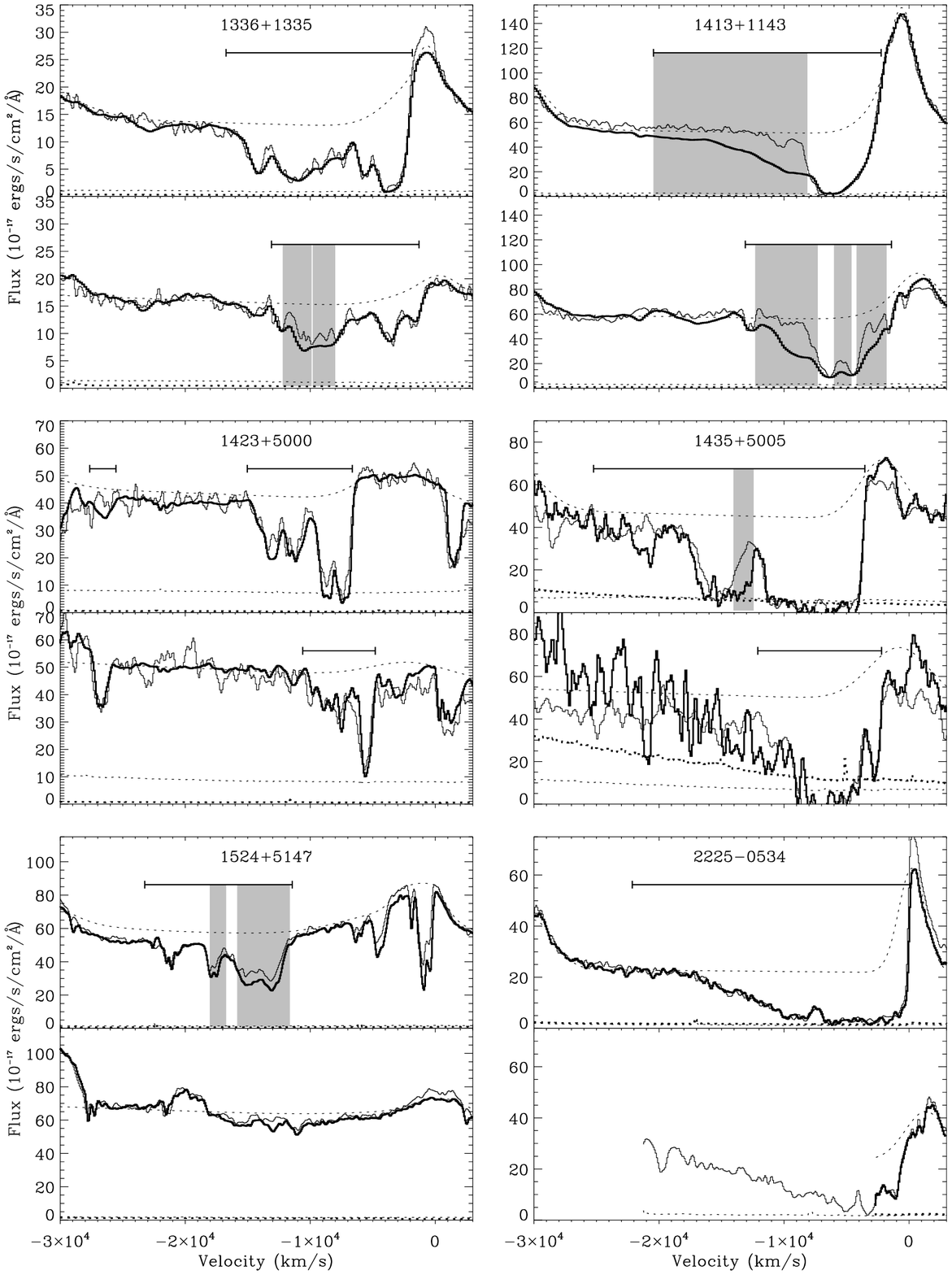}
  \caption{continued...}
\end{figure*}

In Fig. \ref{spectra}, we plot spectra for all 24 objects, showing the long-term comparisons ($\Delta$t = 3.8$-$7.7 yr) between a Lick spectrum and an MDM spectrum. The one exception is 2225-0534, for which we only have short-term Lick data. For each object, we plot the \civ\ absorption region in the top panel and the corresponding \siiv\ absorption region in the bottom panel. The velocity scale is based on the wavelengths of \civ\ and \siiv\ in the observed frame calculated from the redshifts given in Table 1. In order to compare the \civ\ and \siiv\ absorption regions, we use the bluer line in both the \civ\ and \siiv\ doublet for the zero-points of the velocity scales, i.e., 1548.20 \AA\ for \civ\ and 1393.76 \AA\ for \siiv.

We adopt the velocity ranges over which \civ\ BAL absorption occurs defined in Paper 1. These regions were defined based on the definition of BI, i.e. they must contain contiguous absorption that reaches $\geq$10 per cent below the continuum across $\geq$2000 \kms. We apply the same definition when defining the velocity ranges of \siiv\ BAL absorption.

In Paper 1, we defined a pseudo-continuum fit for the fiducial Lick observation used in the long-term analysis for each quasar by fitting a power-law to regions of the spectrum free of absorption and emission. The preferred spectral regions for the fits were 1270$-$1350 \AA\ and 1680$-$1800 \AA, but were adjusted to avoid emission and absorption features as much as possible or due to the limits of the wavelength coverage. We then fit the \civ\ emission lines, using between 1 and 3 Gaussians to define the line profile. In Paper 1, we also fit the \siiv\ emission in cases where \civ\ absorption overlaps with the \siiv\ emission. For this work, we additionally fit the \siiv\ emission in the fiducial Lick observation for all the quasars. For the long-term comparisons below, we fit the \siiv\ emission for the MDM spectrum in cases where the emission line varied. Some special cases where there were difficulties with fitting the \siiv\ emission, such as 1011+0906 and 1309$-$0536, are discussed further in Section 3.3 below.

When comparing multiple epochs, we scaled all the spectra to the fiducial Lick spectrum used for the pseudo-continuum fit. We only fit the power-law continuum to one spectrum for each object, so any errors in this continuum fit will not effect our main variability results. To match the individual epochs for each quasar, we adopt a simple vertical scaling that matches the spectra along the continuum redwards of the \civ\ emission line (i.e. from 1560 \AA\ to the limit of the wavelength coverage), between the \siiv\ and \civ\ emission ($\sim$1425-1515 \AA), and between the Ly$\alpha$ and the \siiv\ emission ($\sim$1305-1315 \AA). For the few cases where a simple scaling did not produce a good match and there were disparities in the overall spectral shape between the comparison spectra, we fit either a linear function (for 0903+1734, 1413+1143, 1423+5000, and 1524+5147) or quadratic function (for 0019+0107 and 1309-0536) to the ratio of the two spectra across regions that avoid the BALs. We then multiplied this function by the SDSS or MDM spectrum to match the fiducial Lick data.

With the spectra for each object matched, we used visual inspection to identify velocity intervals with a width of at least 1200 \kms\ that varied. We identify intervals of variability separately for \civ\ and \siiv\ absorption. We then calculate the average flux and associated error for each candidate variable interval in each of the two epochs being compared. The error on the average flux is given by
\[
    \sigma_{f}^{2} = \frac{1}{n^{2}} \sum_{i=1}^{n} \sigma_{i}^{2}
\]
where $\sigma_{i}$ is the error on an individual pixel, taken from the error arrays as displayed in Fig. 1, and $n$ is the number of pixels in the candidate variable interval. We then calculate the flux difference between the two spectra and place an error on this flux difference using the error on the average flux from each epoch. We include all intervals of variability where the flux differences are at least 4$\sigma$. Any interval which varied by at least 4$\sigma$ was readily identified by our initial inspection procedure.

However, photon statistics alone are not sufficient for defining real variability, so we took a conservative approach, described in more detail in Paper 1, whereby we omit ambiguous cases of variability, even if they meet the 4$\sigma$ threshold. Flux calibrations, a poorly constrained continuum placement, and underlying emission-line variability can all add additional uncertainty to identifying variability. For example, in 1011+0906 and 1309$-$0536, there might be BAL absorption, and variability, on top of the \siiv\ emission line (see Fig. 1). However, it is too ambiguous to be included in this study. See Paper 1 for further examples of intervals of potential variability that were not included because of additional uncertainties and Section 3.3 below, where we comment further on certain individual quasars. We include narrow intervals of variability such as the shaded region in \siiv\ in 0043+0048 and the shaded region in \civ\ in 1246$-$0542 in Fig. 1, where the flux differences are 6.3$\sigma$ and 5.6$\sigma$, respectively. These regions meet the aforementioned thresholds, and the comparison spectra match well in regions of the continuum free of emission and absorption and on either side of the variability interval. We also include regions such as those shaded in 1011+0906 in Fig. 1 because even though the errors are slightly higher in the MDM spectra shown, the flux differences are still 7.7$\sigma$ for the variable region in \civ\ and 6.3$\sigma$ and 7.9$\sigma$ for the two regions of variability in \siiv. Overall our approach is designed to be conservative; we try to exclude marginal cases of variability to avoid overestimating the true variability fractions.

We calculated the absorption strength, $A$, of the BALs in our sample, where $A$ is the fraction of the normalized continuum flux removed by absorption (0 $\leq$ A $\leq$ 1) within a specified velocity interval. These calculations are described in Paper 1. Very briefly, we divide each interval of variability and absorption, as defined above, into equal-sized bins of width 1000 to 2000 \kms, with the final bin size depending on the total velocity width of the specified interval. Then, for each quasar, we adopt the same bin size for the epochs being compared and calculate  $\langle A\rangle$ and $\Delta A$ in every individual bin.

One of the difficulties in directly comparing \civ\ to \siiv\ absorption is the wider doublet separation in \siiv\ (500 \kms\ in \civ\ versus 1900 \kms\ in \siiv). This can cause the \siiv\ absorption intervals to be wider than those in \civ. The only effect this should have on the variability results in Section 3 below is that there may be portions of \siiv\ absorption that are detected as variable but the corresponding velocity intervals in \civ\ may be too narrow to pass our variability threshold. We comment further on this in Section 3.1. We mark the regions defined as BAL absorption (horizontal bars) and variability (shaded rectangles) in Fig. \ref{spectra}. The absorption and variability regions for \civ\ were defined in Paper 1. We defined the absorption and variability regions for \siiv\ independently from what we found for \civ. See Section 3.1 for a full discussion of these results.

\begin{figure}
  \includegraphics[width=84mm]{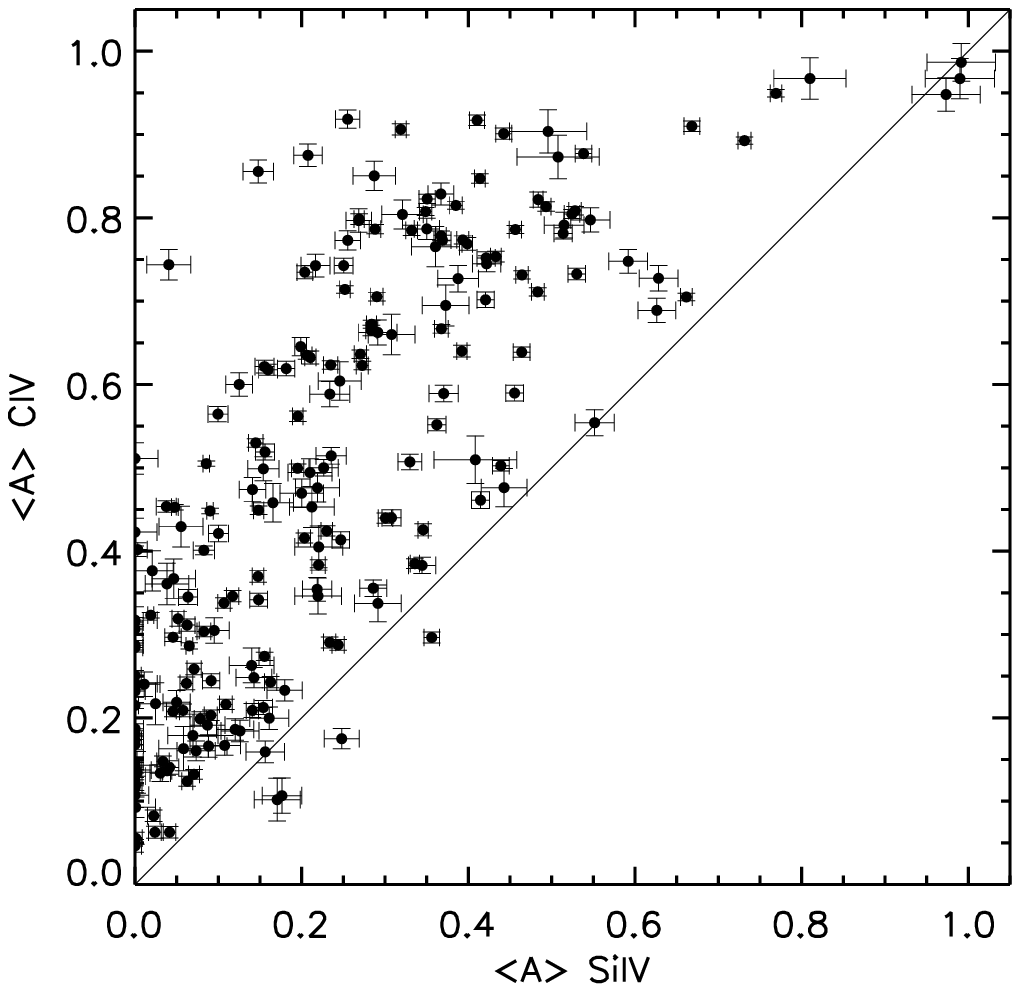}
  \caption{The average normalized absorption strength, $\langle A\rangle$, in \civ\ versus the
    $\langle A\rangle$ in \siiv\ for each absorption bin in each quasar in the long-term subsample.
    The error bars represent the 1$\sigma$ errors based on photon statistics.}
  \label{AvA}
\end{figure}

In Fig. \ref{AvA}, we plot the relationship between absorption strength, $\langle A\rangle$, in \civ\ and $\langle A\rangle$ in \siiv\ for the long-term spectra shown in Fig. 1. Each point represents a different absorption bin in an individual quasar, such that individual quasars contribute multiple points to this plot. The diagonal line through the plot represents equal strength in both lines. Since we fix the velocity scale based on the bluer doublet member in \civ\ and in \siiv\ and the doublet separation is wider in \siiv, the edges of the \siiv\ absorption troughs tend to extend redward of the edges of the corresponding \civ\ troughs (see, for example, 0302+1705 in Fig. 1). As mentioned above, to calculate the $\langle A\rangle$ values, we divide each BAL into bins. We thus remove the redmost bin for each absorption trough because \siiv\ may have a greater strength than \civ\ in those bins due to the wider doublet separation and not necessarily because the \siiv\ absorption is actually stronger than \civ. We also plot 1$\sigma$ error bars, calculated by using the error spectra shown in Fig. 1 and averaging over the velocity interval for each bin. We point out that non-statistical errors, e.g., from the continuum fitting, can increase the error in $\langle A\rangle$ measurements by up to 0.05 to 0.1. It is clear from Fig. \ref{AvA} that absorption in \civ\ is roughly as strong or stronger than the corresponding \siiv\ absorption. In some cases, the \civ\ has no detectable corresponding \siiv\ absorption at all.

There are almost no well-measured cases of \civ\ absorption weaker than \siiv.  We did find a few absorption bins where \siiv\ appeared to be stronger than \civ, but some of these bins might not actually have absorption due to \siiv, or due solely to \siiv. In about 10\% of BAL spectra, there are lower ionization lines, such as \cii\ $\lambda$1335 and \al\ $\lambda$1855,1863 \citep{Trump06}. The wavelength of \cii\ places any \cii\ absorption line at a velocity of $<$$-$13 500 \kms\ in \siiv\ velocity space. In order to look for interloping \cii\ lines, we checked all of our objects for another low ionization species, \al, which exists redward of \civ, in a region of the spectrum relatively uncontaminated by other lines. We found that in two cases (1232+1325, 1331$-$0108), the velocity of the \al\ line puts the corresponding \cii\ line within a \siiv\ BAL trough. We therefore removed the bins affected by \cii\ from Fig. \ref{AvA}, and we exclude the contaminated \siiv\ regions in these two quasars from the analysis below. There are still a few points below the one-to-one line in Fig. \ref{AvA}. However, these few points are mostly within 3$\sigma$ of the one-to-one line and therefore are consistent with equal strength changes in \civ\ and \siiv. The one point that is just beyond 3$\sigma$ from the line corresponds to the interval $-$3100 to $-$1200 \kms\ in 1303+3048.

\section[]{Results}

\subsection{Variability in \siiv\ versus \civ\ BALs}

In this section, we directly compare the variability of \siiv\ to \civ\ in the ``long-term" dataset from Paper 1. This involves 2 epochs of data for 23 quasars separated by 3.8 to 7.7 yrs. Our main goal is to discriminate between the possible causes of BAL variability. We begin by looking at what fractions of quasars exhibited \civ\ and \siiv\ variability to determine if \siiv\ varies more or less often than \civ. In Paper 1, we found a correlation between the incidence of \civ\ BAL variability and outflow velocity and absorption strength. Here we investigate whether similar trends exists for \siiv\ BALs. We then look at the relationship between the incidence of \civ\ variability and \siiv\ strength. Last, we compare the change in strength for the two lines when they both vary.

In the long-term dataset in Paper 1, 15 out of 23 quasars (65 per cent) exhibited \civ\ BAL variability and 11 out of 19 (58 per cent) exhibited \siiv\ BAL variability, at any measured velocity. We do not have data covering the \siiv\ region for 2 of our quasars, and another 2 quasars do not have \siiv\ BALs. This comparison between \civ\ and \siiv\ is complicated because the absorption in \civ\ is not always accompanied by corresponding absorption (e.g., at the same velocities) in \siiv\ (Figs. 1 and 2). In addition, we are observationally less sensitive to absorption and variability at high velocities in \siiv, compared to \civ, because those wavelengths can have poorer signal-to-noise ratios and larger uncertainties in the continuum placement caused by blends with underlying broad emission lines. Altogether this means we are more sensitive to variability in \civ\ than \siiv\ in our data set.

To make a fair comparison between the incidence of variability in the two lines, we recalculate the above fractions while considering just the flow speeds at $>$$-$20 000 \kms. We adopt this velocity as the cutoff because, in some spectra, there is emission due to \oi\ at $\sim$$-$20 500 \kms\ in the \siiv\ absorption region (see also \citealt{Gibson10}). With this additional restriction, we find that \siiv\ is more likely to vary than \civ. In particular, 35 per cent (8/23) of quasars exhibited \civ\ variability and 47 per cent (9/19) exhibited \siiv\ variability. The dramatic reduction in the \civ\ variability recorded this way, compared to the 65 per cent quoted above, is due to i) consideration of a narrower velocity range, and ii) the specific exclusion of high velocities, which are the most likely to show variability (Paper 1). Nearly half of the occurrences of \civ\ variability detected in our data set are at high velocities, i.e., v $<$$-$20 000 km/s. The incidence of \civ\ variability further reduces to 31 per cent (6/19) if we only include the 19 quasars which have complete spectral coverage across \siiv\ and have a \siiv\ BAL. The further decline in the \civ\ variability in this case probably occurs because the two quasars excluded for having no \siiv\ BAL have weak \civ\ lines, and weak \civ\ lines are more likely to vary than strong ones (Paper 1). Thus, we again removed \civ\ BALs that are more likely to vary.

Overall, it is important to realize that a number of factors can affect the measured incidence of BAL variability. Our comparisons show that, over matching velocity ranges, \siiv\ BALs have a significantly higher incidence of variability than \civ\ BALs. This difference is probably related to the different line strengths. In particular, the \siiv\ BALs are generally weaker than \civ\ BALs (Fig. \ref{AvA}), and weaker lines tend to be more variable (Paper 1 and Figs. \ref{hist_siv} and \ref{histA} below).

To more directly compare \siiv\ and \civ\ BAL variability, we examine the individual velocity intervals over which the variability occurs. We consider intervals at all velocities here and in the remainder of this section. Variability in \civ\ occurred in a total of 20 velocity intervals. Ten of these intervals have measurable \siiv\ absorption. Nine of these 10 intervals, or 90 per cent, showed \siiv\ variations in the same sense (either getting stronger or weaker) as the \civ\ changes. There is only one interval (in 0119+0310) that exhibits variability in \civ, without corresponding variability at the same velocities in \siiv. Conversely, we find long-term \siiv\ BAL variability in a total of 22 velocity intervals. All of the variable \siiv\ intervals have significant corresponding \civ\ absorption, and 10 of these intervals showed \civ\ variations in the same sense as the \siiv\ (45 per cent).

As mentioned in Section 2.2, \siiv\ has a wider doublet separation than \civ, causing some of the \siiv\ absorption and variability intervals to be wider than the corresponding \civ\ intervals. For the intervals of \siiv\ variability without corresponding \civ\ variability, we looked for any evidence of variability in \civ\ that was not included because the width of the candidate varying region was too narrow to meet our variability threshold (see Section 2.2). There is only one case where we detect a marginal narrow variability region in \civ\ corresponding to a region in \siiv\ classified as variable (in 0903$+$1734). Even if we were to count this as a variable \civ\ interval, still only 50 per cent of \siiv\ variability intervals would have corresponding \civ\ variability. Therefore, while 91 per cent of the intervals of \civ\ variability had corresponding \siiv\ variability, only $\sim$50 per cent of the intervals of \siiv\ variability had corresponding \civ\ variability. These results reinforce our main conclusion above, that \siiv\ BALs are more likely to vary than their \civ\ counterparts.

\begin{figure}
  \includegraphics[width=84mm]{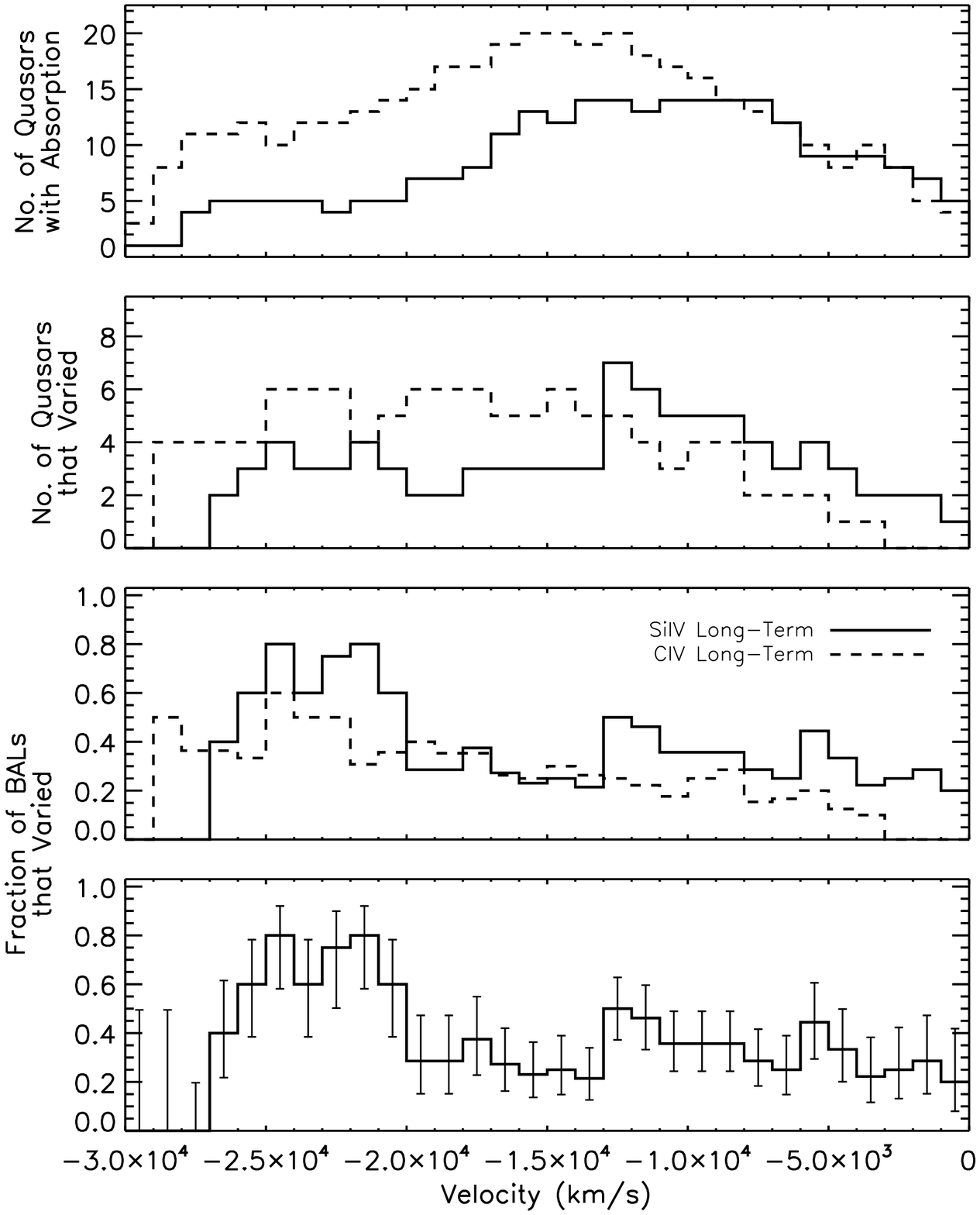}
  \caption{The top two panels show the number of occurrences of \siiv\ (solid lines) and \civ\
    (dashed lines) BAL absorption and variable absorption versus velocity. The third panel is the
    second panel divided by the first. The bottom panel shows the same curve from the third panel
    for \siiv\ with 1$\sigma$ error bars.}
  \label{histV}
\end{figure}
Next we examine the dependence of \siiv\ variability on velocity and absorption strength, matching our analysis of \civ\ BALs in Paper 1. Figure 3 shows the incidence of \siiv\ absorption and \siiv\ variability versus velocity. For comparison, this figure also shows the corresponding data for \civ\ taken directly from fig. 3 in Paper 1 (dashed curves). The top two panels display the number of quasars with \siiv\ BAL absorption and with \siiv\ BAL variability at each velocity (solid curves). The third panel is the second panel divided by the top one, which gives the fraction of \siiv\ BALs that varied at each velocity. The top panel shows clearly that there are more \civ\ BALs than \siiv\ BALs at higher velocities.

In Paper 1, we showed that the incidence of \civ\ variability increases significantly with increasing velocity. This trend is not evident in the \siiv\ data. Figure 3 displays 1$\sigma$ error bars  based on \citet{Wilson27} and \citet{Agresti98}. These errors are based on counting statistics for the number of quasars with absorption and variability at each velocity. We performed a least-squares fit, and the slope of the \siiv\ data is $-6.58 \pm 4.33 \times 10^{-6}$, in the formal unit, fraction per \kms. The slope is non-zero at just a 1.5$\sigma$ significance. Therefore, while there {\it might} be a weak tendency for more variability in \siiv\ at higher velocities (Figure 3), the trend is not statistically significant.

\begin{figure}
  \includegraphics[width=84mm]{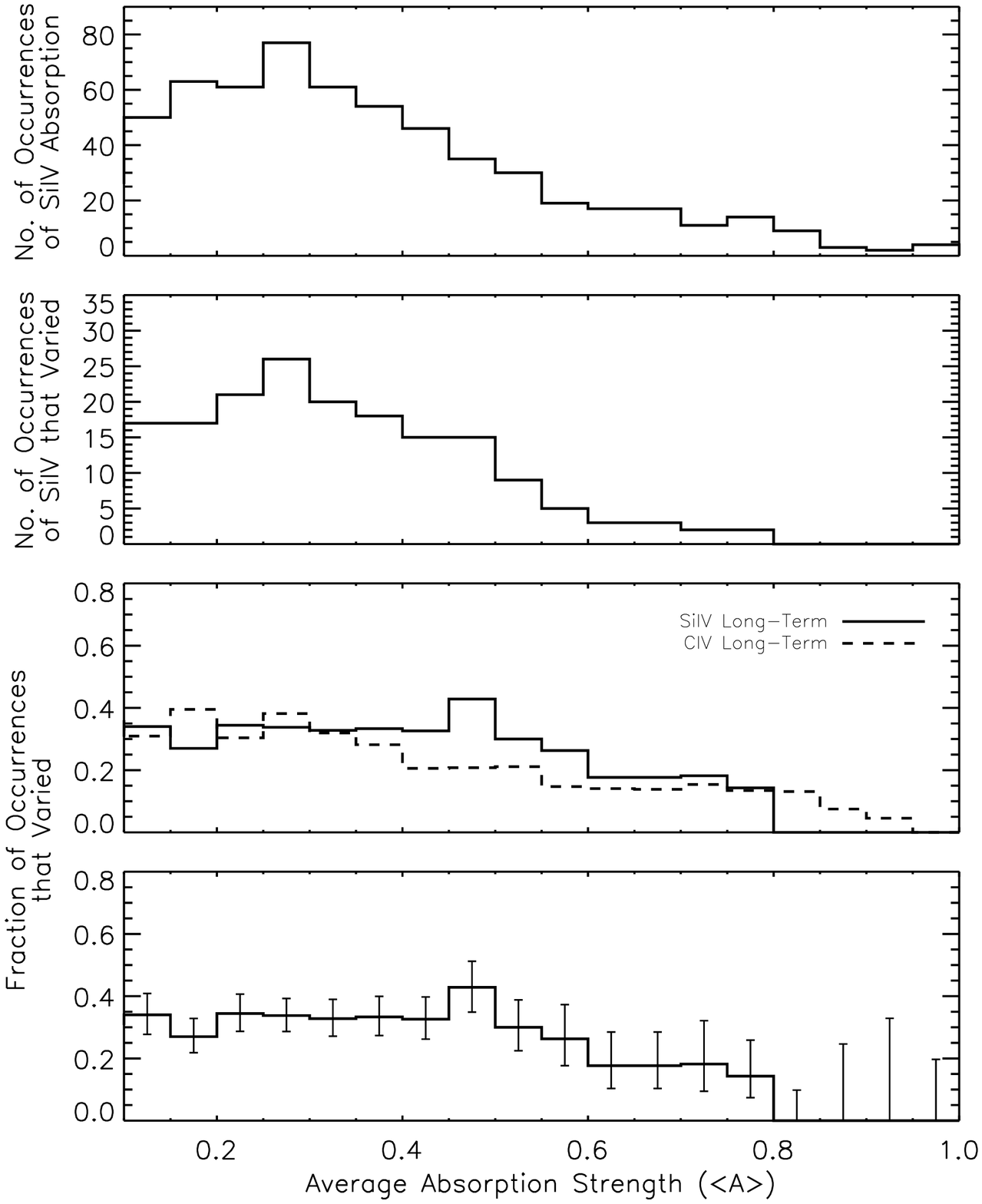}
  \caption{The top two panels show the number of occurrences of \siiv\ BAL absorption and
    variability versus the average normalized absorption strength, $\langle A\rangle$, in \siiv. The
    third panel is the second panel divided by the first. The bottom panel shows the same curve from
    the third panel with 1$\sigma$ error bars overplotted.}
  \label{hist_siv}
\end{figure}

To further match our analysis from Paper 1, we looked at the relationship between the incidence of \siiv\ variability and the absorption strength, $\langle A\rangle$, in \siiv. As described in Paper 1 (and Section 2 above), we divide each BAL into $\sim$1200 \kms\ bins, and we treat each bin as an individual occurrence of absorption. In Fig. \ref{hist_siv}, the top two panels show the number of these occurrences of \siiv\ absorption and the number of these occurrences that varied at each value of \siiv\ absorption strength, $\langle A\rangle$. An individual quasar can contribute more than once to each point in the histogram. The third panel is the second panel divided by the top one, and the bottom panel shows the same curve from the third panel with error bars plotted, calculated in the same way as for Fig. \ref{histV}. We find only a weak trend between \siiv\ variability and \siiv\ strength. The slope of the plot is $-0.315 \pm 0.090$ fraction per unit absorption strength, which is non-zero at a 3.5$\sigma$ significance. This is much weaker than the trend between the incidence of \civ\ variability and \civ\ strength found in Paper 1. We overplot this curve for \civ\ from fig. 5 of Paper 1 in the third panel of Fig. 4 (dashed curve). This indicates that the occurrence of variability in \siiv\ is less sensitive to the strength of the line than the occurrence of variability in \civ.

\begin{figure}
  \includegraphics[width=84mm]{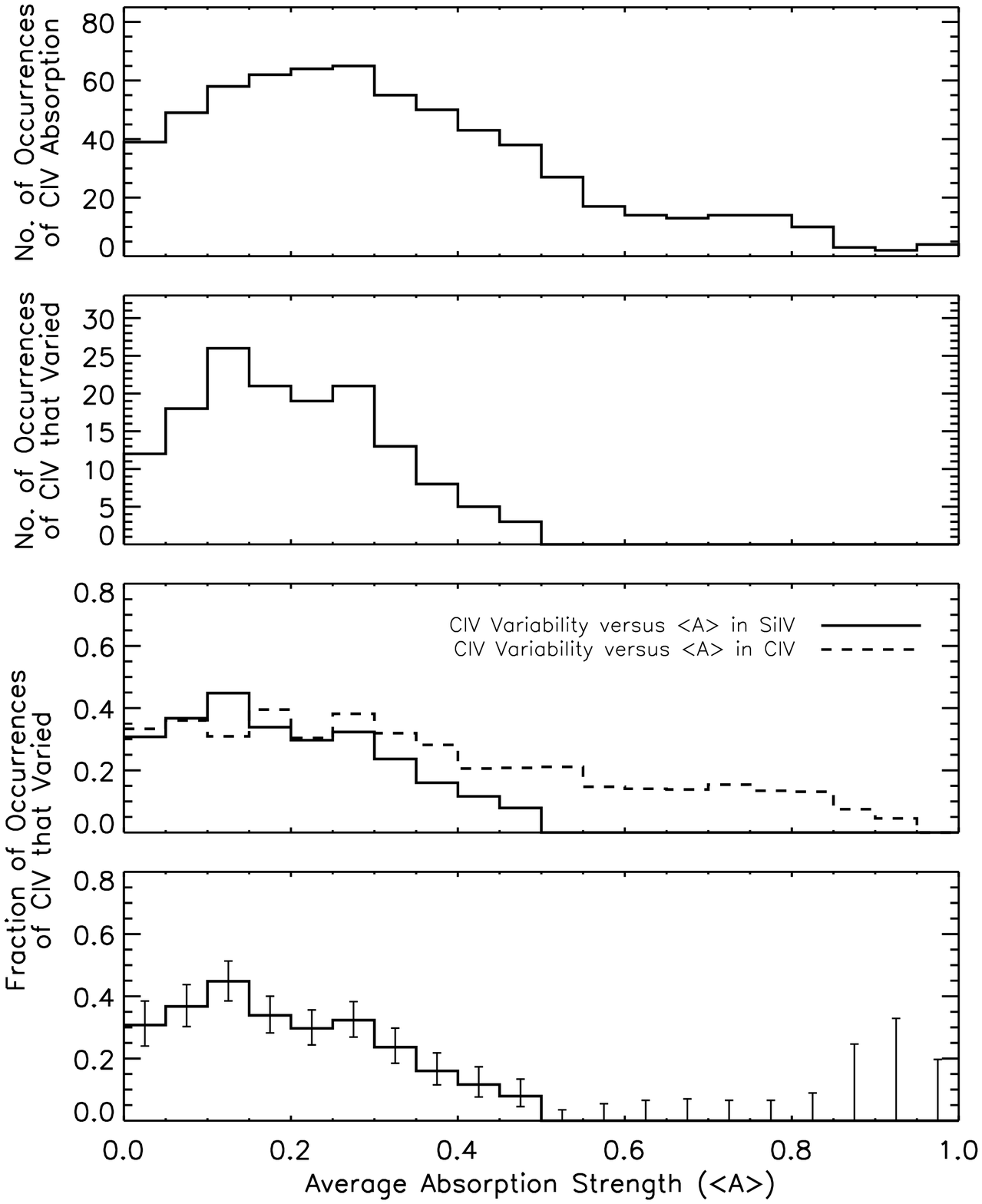}
  \caption{The top two panels show the number of occurrences of \civ\ BAL absorption and
    variability versus the average normalized absorption strength, $\langle A\rangle$, at the
    same velocities in \siiv. The third panel is the second panel divided by the first, and the bottom
    panel shows the same curve from the third panel with 1$\sigma$ error bars. In the third panel, we
    overplot the fraction of occurrences of \civ\ absorption that varied versus $\langle A\rangle$ in
    \civ.}
  \label{histA}
\end{figure}

Next, we looked at the relationship between the incidence of \civ\ BAL variability and the strength of the \siiv\ absorption, $\langle A\rangle$, at the same velocities. Si is known to be less abundant than C, when solar abundances are assumed, and the high ionization typical in BALs favors \civ\ (see Section 5). Therefore, the optical depth in \civ\ is higher than in \siiv, so the stronger the \siiv\ absorption is, the more likely \civ\ is to be saturated. In Fig. \ref{histA}, we plot the number of occurrences of \civ\ absorption that occur at the same velocities in the same spectra as each \siiv\ $\langle A\rangle$ value and then the number of these occurrences that varied in the second panel. As in Fig. \ref{hist_siv}, an individual quasar can contribute more than once to each point in the histogram. The third panel is the second panel divided by the top one, showing the fraction of occurrences of \civ\ absorption that varied at each \siiv\ absorption strength value. As in Figs. \ref{histV} and \ref{hist_siv}, the bottom panel of Fig. \ref{histA} shows the 1$\sigma$ error bars. The slope of these points is $-0.643 \pm 0.066$ fraction per unit absorption strength, which is non-zero at a 10$\sigma$ significance. Fig. \ref{histA} thus indicates that the incidence of \civ\ variability decreases with increasing \siiv\ absorption strength. Therefore, when the \siiv\ absorption is stronger and the \civ\ absorption is more likely to be saturated, the incidence of \civ\ variability decreases. In fact, whenever the \siiv\ absorption strength is greater than 0.5, the corresponding \civ\ absorption at the same velocity does not vary.

As a further comparison to Paper 1, we overplot in the third panel of Fig. \ref{histA} the fraction of occurrences of \civ\ absorption that varied versus \civ\ absorption strength (dashed curve). In Paper 1, we concluded that, for \civ\ BALs, weaker lines are more likely to vary than stronger lines. Fig. \ref{histA} shows that \civ\ lines are even more likely to vary when the corresponding \siiv\ lines are also weak.

\begin{figure}
  \includegraphics[width=84mm]{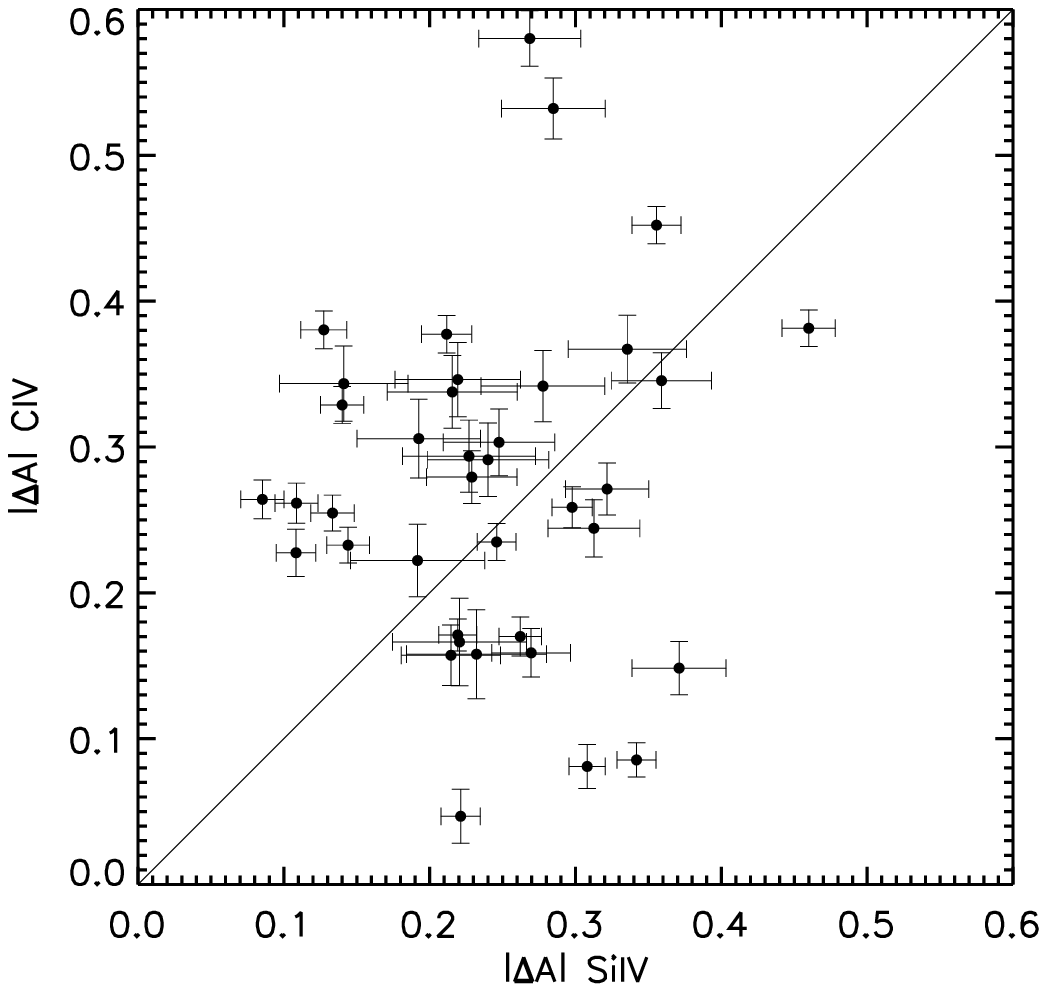}
  \caption{The change in strength of \civ\ BAL absorption versus the change in strength of \siiv\
    BAL absorption in velocity intervals where both lines varied. The error bars are calculated as in
    Fig. \ref{AvA}.}
  \label{dAvdA}
\end{figure}

We also investigate the change in strength, $\vert\Delta A\vert$, in \civ\ compared to the corresponding changes in \siiv\ in the same velocity interval (Fig. \ref{dAvdA}). There are 6 quasars in which the velocity intervals of \civ\ and \siiv\ variability overlap, and we only compare the velocity intervals where {\it both} lines varied. Each point represents one bin in one of these quasars, as described above for Fig. \ref{histA}. The diagonal line through this plot corresponds to equal strength changes in both lines. There is clearly no correlation between the strength changes in \civ\ versus \siiv\ evident in this figure. Despite \siiv\ varying more often than \civ, the strength changes in \siiv\ are not always greater than in \civ.

\begin{figure}
  \includegraphics[width=84mm]{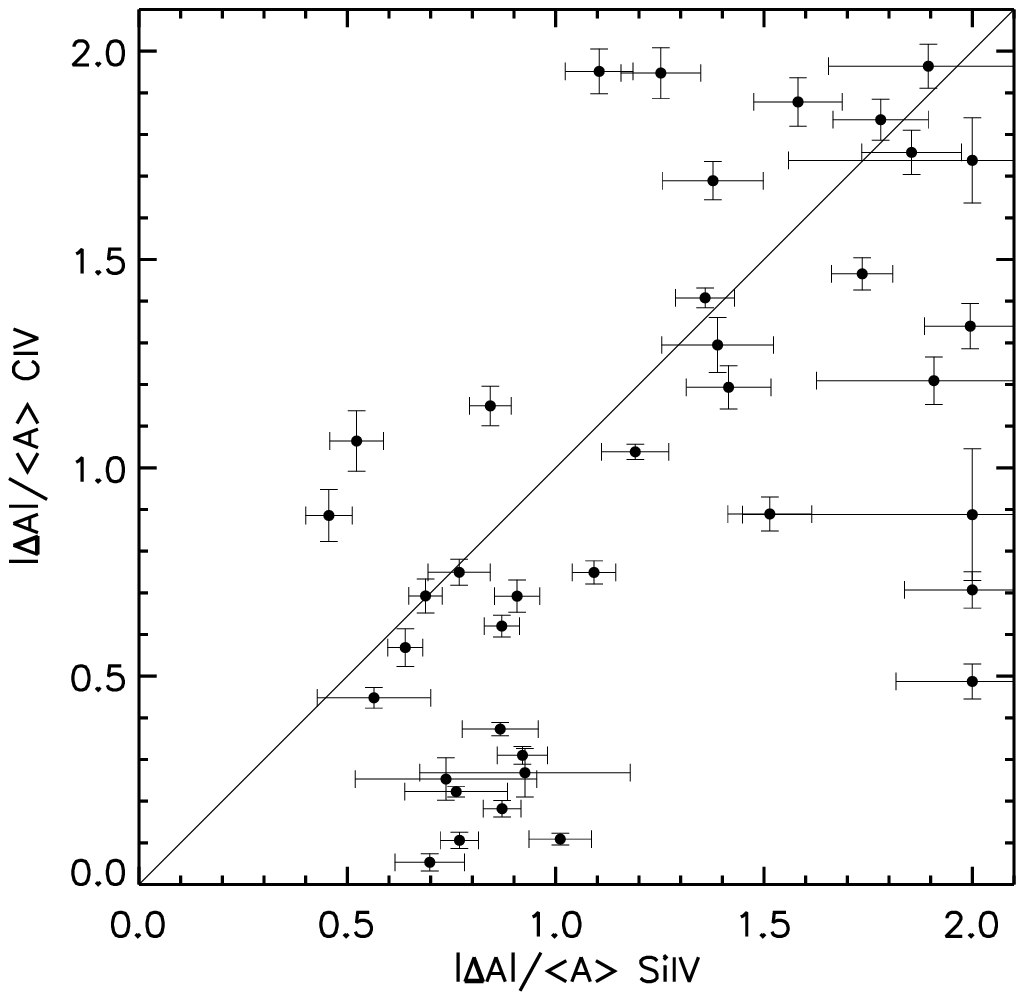}
  \caption{The fractional change in strength of \civ\ BAL absorption versus the fractional change in
    strength of \siiv\ BAL absorption in velocity intervals where both lines varied. The error bars are
    calculated as in Fig. \ref{AvA}.}
  \label{AAvAA}
\end{figure}

Finally, Fig. \ref{AAvAA} shows the fractional change in strength, $|\Delta A|/\langle A\rangle$, in \civ\ compared to $|\Delta A|/\langle A\rangle$ in \siiv, again for velocity intervals where both lines varied. As in Fig. \ref{dAvdA}, there are points lying above the line, representing greater strength changes in \civ\ than in \siiv. However, there is a weak trend towards greater fractional change in strength in \siiv, which is consistent with the correlation found between fractional change in EW in \civ\ and \siiv\ in \citet{Gibson10}.

As mentioned in Section 2.1, we include two quasars that have BI=0 because they do have broad absorption, but it falls outside the velocity range, $-$25 000 to $-$3 000 \kms, used in the strict definition of BI. We also include broad absorption in other quasars in our sample that falls outside of this velocity range. Their inclusion has minimal impact on our results because most of the \siiv\ broad absorption in our sample falls within this velocity range. For the quasars with BI=0, 0846+1540 does not contain any \siiv\ broad absorption at all, and 0302+1705 contains broad absorption at low-velocity, which did not vary in either \civ\ or \siiv.

We can summarize our comparisons between the \civ\ and \siiv\ BAL variabilities as follows: 1) \siiv\ BALs are more likely to vary than \civ\ BALs. The fractions of quasars showing variability in our long-term 2-epoch data set are 31\% in \civ\ and 47\% in \siiv\ if we consider only the well-measured velocity range $v > -$20 000 km/s and include only quasars with both \siiv\ and \civ\ BALs detected. 2) The variabilities usually occur in just portions of the BAL troughs. 3) When changes occur in both \siiv\ and \civ, they always occur in the same sense (i.e., with both lines getting either weaker or stronger). They also occur in overlapping but not necessarily identical velocity ranges. 4) The trend for a higher incidence of \civ\ variability at higher velocities, which we reported in Paper 1, is not clearly evident in the \siiv\ data. Finally, 5) there is no correlation between absorption strength changes in \civ\ versus \siiv\ when they both vary; although, there is a weak trend towards greater fractional change in strength in \siiv.

\subsection{Multi-Epoch Monitoring of BALQSOs}

We now expand our analysis to the full dataset, which includes 2 to 10 epochs of data for each object (Table 1). Including all of these epochs and considering all measured velocities, the fraction of quasars that showed \civ\ BAL variability is 83 per cent. This is a significant increase from the 65 per cent we derived considering only two long-term epochs, or the 39 per cent derived from only two short-term epochs (Section 3.1 and Paper 1). Clearly, including more epochs of data increases the observed variability fractions. Moreover, these larger variability fractions apply to roughly the same time-frame as our 2-epoch long-term data set. Therefore, the multi-epoch data did not find new occurrences of variability at some other time; they identified variability missed by the 2-epoch measurements.

To investigate the multi-epoch behaviours of BAL variability, we compared all the spectra obtained for each quasar. From one object to another, there are large differences in the widths of the varying regions and the amplitudes of the changes (e.g., see Figs. 1 and \ref{dAvdA}). However, there are certain general trends that most, if not all, the quasars follow. In particular, the variability almost always occurred within just a portion of a BAL and not in the entire trough. In nearly all the quasars, the variability occurred over the same velocity interval(s) between each epoch. Finally, when there are multiple velocity intervals of variability within the same quasar, the changes in these separate intervals almost always occur in the same sense. Similarly, as described in Section 3.1, when there is variability in both \civ\ and \siiv, they also vary in the same sense.

We also find no clear evidence for velocity shifts that would be indicative of acceleration or deceleration in the flows. The constraints on velocity shifts are difficult to quantify in BALs because there can be complex profile variabilities, but we specifically search for and did not find cases where a distinct absorption feature preserved its identity while shifting in velocity. Despite the large outflow velocities, there is no clear evidence to date for acceleration or deceleration in BALs, or in any other outflow lines (i.e. NALs and mini-BALs; e.g. \citealt{RodriguezH11}).

\begin{figure*}
  \includegraphics[width=164mm]{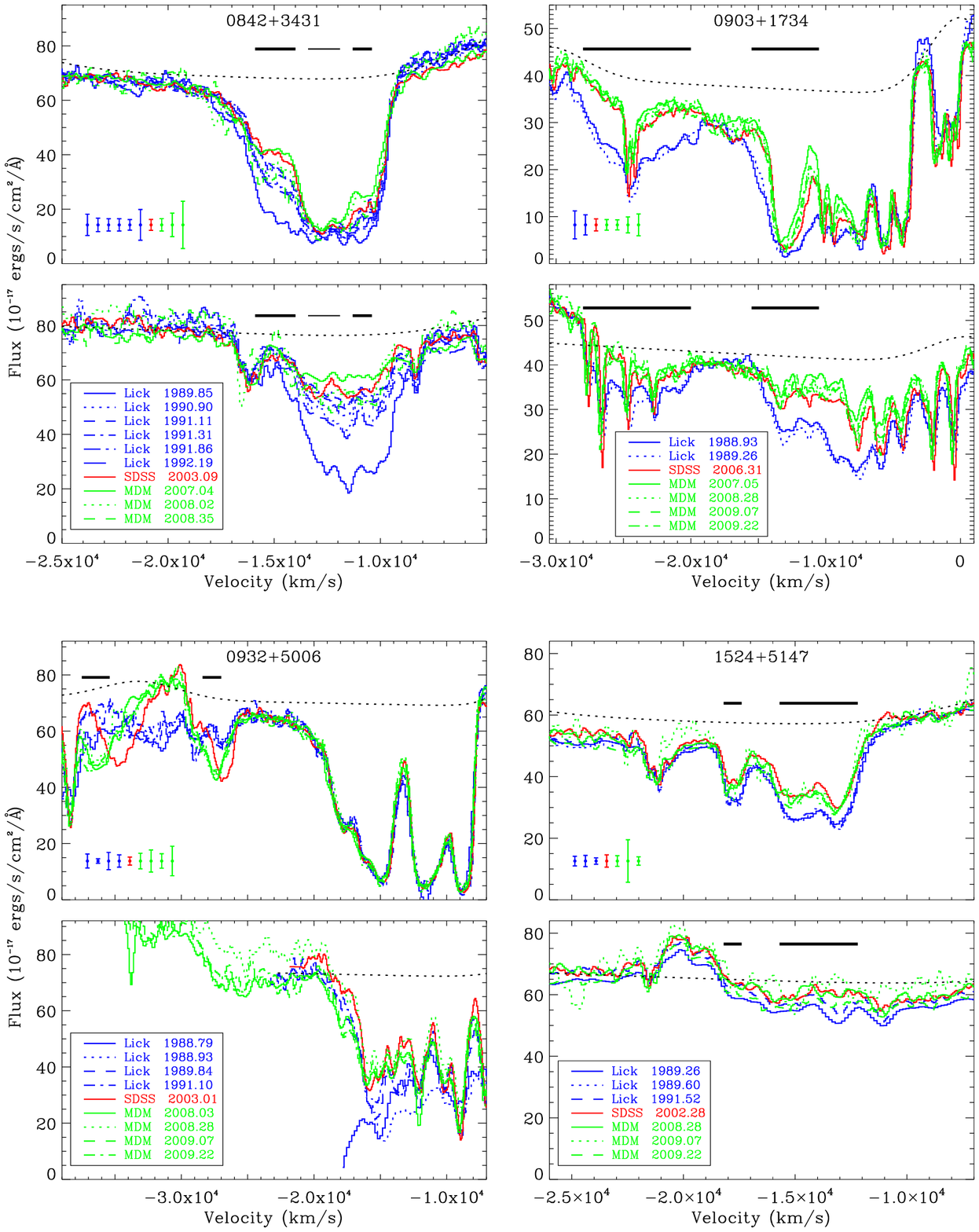}
  \caption{Spectra of the \civ\ (top panel) and \siiv\ (bottom panel) BALs in four well-sampled
    quasars from our sample, after smoothing three times with a binomial function. The blue curves
    are the Lick spectra, red curves are SDSS spectra, and green curves are MDM spectra. The bold
    bars mark varying regions identified for \civ. The average error for each spectrum is shown in
    the top panel for each quasar, where the height of the error bar represents $\pm1\sigma$.}
  \label{multi}
\end{figure*}

We highlight below a few well-sampled cases to illustrate these general trends in the data. Fig. \ref{multi} shows specifically the quasars for which we have at least 7 measured epochs including one from the SDSS, which helps to span the time gap between the early Lick data and our recent MDM observations. For each object, the top panel shows the \civ\ BAL(s) and the bottom panel shows the \siiv\ BAL(s). The blue curves show the early Lick data, the red curves show the intermediate SDSS data, and the green curves show the MDM spectra. We note that in 1524+5147 there is an \oi\ emission line centered at $\sim$$-$20 500 \kms\ in the \siiv\ panel.

\begin{figure*}
  \includegraphics[width=170mm]{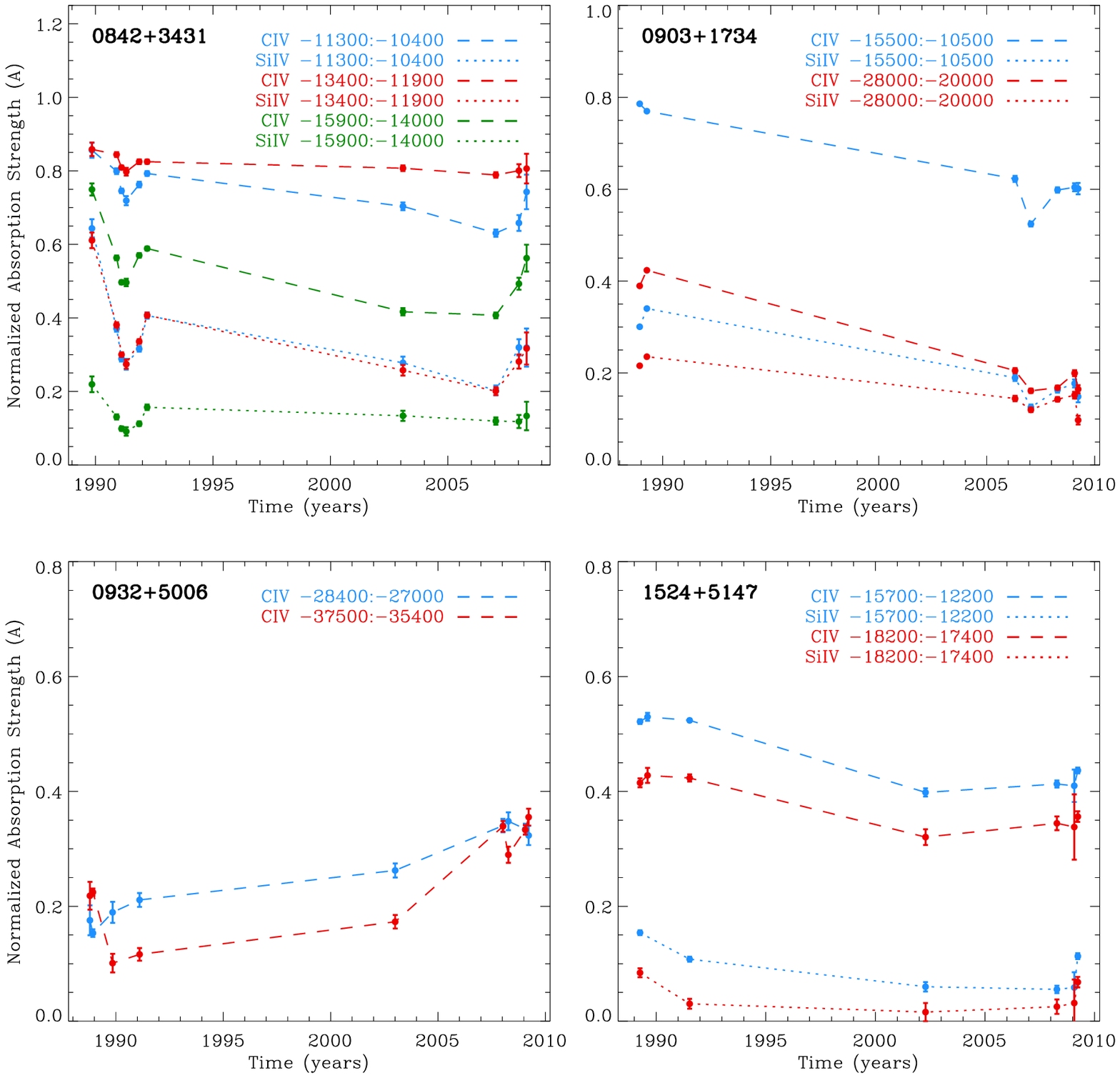}
  \caption{The absorption strength in both \civ\ (dashed lines) and \siiv\ (dotted lines) as a function
    of time in the velocity intervals indicated by bars in Fig. \ref{multi}.}
  \label{AvT}
\end{figure*}

The bold bars in Fig. \ref{multi} mark intervals of variability identified for \civ. The velocity ranges are guided by the intervals defined for \civ\ in Paper 1 and are adjusted to cover the core of the varying region and avoid the edges where the variability is less pronounced. These same velocity ranges are applied to all the epochs plotted here and to \siiv\ for comparison. The one exception is the thin bar marking a region of variability in \siiv\ in 0842+3431 with minimal corresponding variability in \civ. For all of these varying regions, we calculate the absorption strength, $A$, for the defined velocity intervals in each epoch and then plot these $A$ values versus time in Fig. \ref{AvT}.

We plot $A$ instead of EW, in Fig. \ref{AvT}, in order to highlight the intervals that varied. Using EW would dilute these changes in strength. The different colors correspond to different velocity intervals. The dashed and dotted lines represent changes in \civ\ and \siiv, respectively. We note that these lines do not represent how the $A$ values changed between epochs. They simply connect the measurements from different epochs in order to aide the eye.

In Fig. \ref{multi}, the spectra of 0842+3431 (in the \civ\ BAL), 0903+1734, and 1524+5147 show clearly how portions of BALs can vary. In 0842+3431, there is significant variability in both the blue side and the red side of the \civ\ BAL; although in \siiv, the entire BAL varies (see Section \ref{0842} below). In 0903+1734, there is significant variability at higher outflow velocities, but at lower velocities, there is no variability. This is consistent with the result from Paper 1 that there is a higher incidence of variability at higher velocities. The variable regions in 1524+5147 cover most, but not all, of the BAL. In contrast to the general trends in Paper 1, the blue-most portion of the BAL did not vary. In 0932+5006, however, the entire \civ\ BALs vary at the highest velocities. In terms of \civ\ to \siiv\ comparisons, 1524+5147 is a case where there is weak corresponding absorption, and variability, in \siiv, but no \siiv\ BAL. And, 0932+5006 shows clearly a case where a \siiv\ BAL varied, but the \civ\ BAL did not.

The plots in Fig. \ref{AvT} for 0842+3431, 0903+1734, and 0932+5006 show how the change in $A$ is not always monotonic. The same BAL can grow deeper from one epoch to another, then become shallower again. This is consistent with the results of \citet{Gibson10}. Furthermore, the change in the $A$ value from one epoch to another generally occurs in the same direction (either positive or negative) in both \civ\ and \siiv, which is consistent with the results of Section 3.1. In 0842+3431, 0903+1734, and 1524+5147, where there are two separate intervals of \civ\ variability, the change in strength occurs in the same direction for both intervals. The high-velocity BALs in 0932+5006 vary in concert starting with the 1989.84 epoch through 2008.03. At the earliest and latest epochs, they do not clearly vary in concert, but the changes in $A$ are small and could be affected by changes in the underlying \siiv\ emission line. We comment further on these BALs in 0932+5006 in Section \ref{0932}.

\subsection{Notes on Individual Quasars}

In this section, we comment on individual quasars that are cases of special scientific interest. We also comment on cases where there were specific issues in the analysis or measurements that result in larger uncertainties.

\subsubsection{0119+0310}
\label{0119}

0119+0310 is the only quasar for which we record \civ\ BAL variations without corresponding changes in \siiv\ in our long-term sample (Fig. 1 and Section 3.1). However, these results are very tentative because the \siiv\ absorption is poorly measured across the velocities that varied in \civ. The two long-term spectra for this object, plotted in Fig. 1, have a lower signal-to-noise level than most of the other data in our sample. Furthermore, the \siiv\ absorption at the velocity of \civ\ variability ($\sim$$-$7500 \kms) is very weak, and if the continuum fit is off by even $\sim$5 per cent, this region in \siiv\ might not be considered part of the \siiv\ BAL. If this region is not part of the BAL, then we would not include it in the comparison of \civ\ to \siiv. Therefore, while we have several well-measured cases of \siiv\ variability without corresponding \civ\ variability, we only have this one poorly-measured case of \civ\ variability with no corresponding \siiv\ variability.

This quasar also appears to differ from most of the other quasars in the sample in that the different regions of \civ\ variability do not vary in the same sense (Fig. 1). The two higher velocity variable regions both increase in strength between the Lick and MDM observations, while the lower velocity variable region decreases in strength. As mentioned above, this is one of our least well-measured quasars, so this is a tentative result. We find just two other cases where two regions of \civ\ variability vary in opposite directions (0146+0142 and 1423+5000).

\subsubsection{0146+0142}
\label{0146}

\begin{figure}
  \includegraphics[width=84mm]{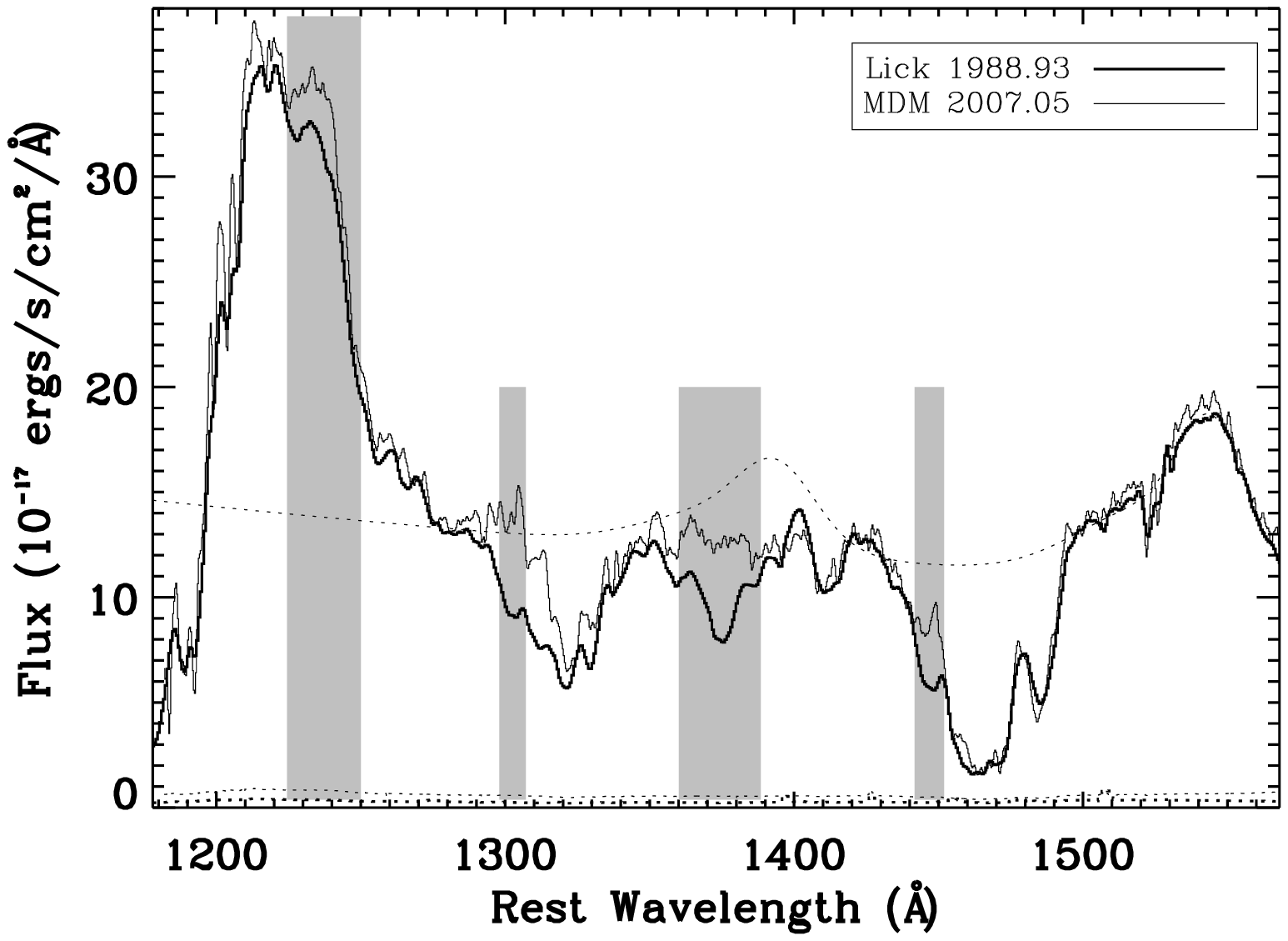}
  \caption{The two long-term epochs for 0146+0142, showing the full spectrum from the
    Ly$\alpha$ emission through the \civ\ emission. The shading here differs from Fig. 1, with the
    right-most shaded regions marking the \civ\ variability and the left-most shaded regions
    showing the corresponding velocities in \siiv\ (but not the exact regions of \siiv\ variability).
    This figure shows evidence of \siiv\ BAL variability on top of the Ly$\alpha$ emission line.}
  \label{0146LT}
\end{figure}

\begin{figure}
  \includegraphics[width=84mm]{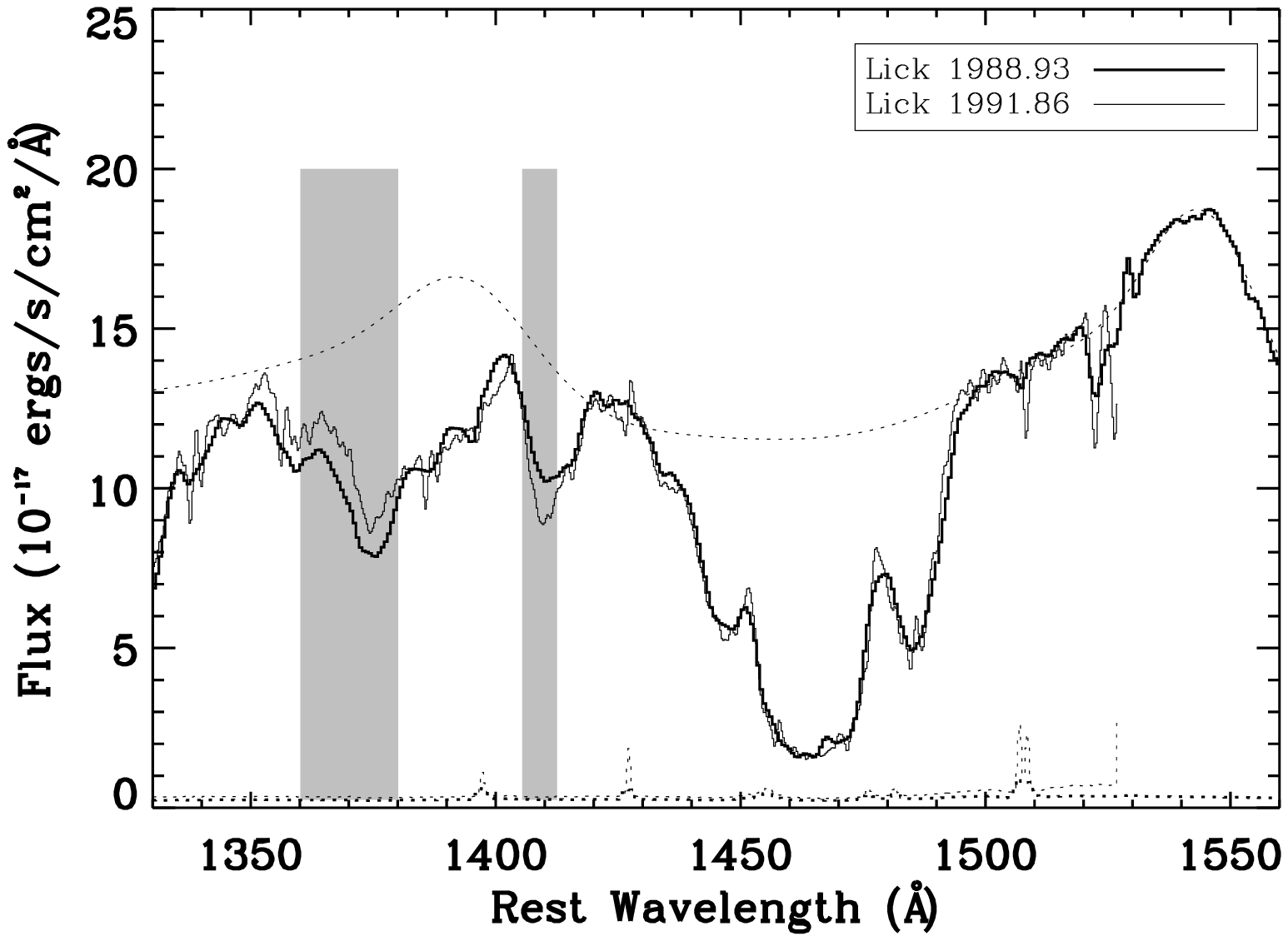}
  \caption{The two short-term epochs for 0146+0142, with the two shaded regions marking \civ\
    variability. The absorption in these two regions vary in opposite directions, with one region
    becoming weaker while the other becomes stronger.}
  \label{0146ST}
\end{figure}

We first note that this object has a high-velocity \civ\ BAL, so the BAL that appears just redward of the marked \siiv\ BAL in Fig. 1 is actually \civ. As mentioned in Paper 1, we can confirm that this is high-velocity \civ, and not \siiv, because if it were \siiv\ absorption, we should see corresponding low-velocity \civ\ absorption (see also fig. 1, \citealt{Korista93}). We also have further confirmation that this is high-velocity \civ\ absorption because we find evidence of corresponding high-velocity \siiv\ absorption on top of the Ly$\alpha$ emission line. Fig. \ref{0146LT} shows the two long-term spectra for 0146+0142 with the two right-most shaded regions marking the \civ\ variability and the two left-most shaded regions marking the corresponding velocities (but not necessarily the entire variable regions) in \siiv.

Another interesting note about 0146+0142 is that in our short-term data, there are two separate regions of \civ\ BAL variability, but they do not vary in the same sense. In Fig. \ref{0146ST}, we plot the two short-term epochs with these two variable regions shaded. The redder varying interval at $\sim$$-$28 300 \kms\ increases in strength, while the bluer interval at $\sim$$-$36 600 \kms\ decreases in strength. As mentioned in Section \ref{0119}, this is one of just three cases showing this behavior.

\subsubsection{0842+3431}
\label{0842}

In 0842+3431, there is significant variability in two distinct regions of the \civ\ BAL trough (marked by bold bars in Fig. \ref{multi}), while the entire \siiv\ trough varies. However, the bluer, and more variable, region in the \civ\ trough (the left-most bold bar in Fig. \ref{multi}) corresponds to a region of weak absorption and variability in \siiv. Some of the variability in the redmost portion of the \siiv\ trough (the right-most bold bar in Fig. \ref{multi}) may be connected with the variability in the red side of the \civ\ BAL trough, but the variability in the core of the \siiv\ trough, marked by the thin bar in Fig. \ref{multi}, cannot be explained by the wider \siiv\ doublet separation alone. Some of the \siiv\ variability is occurring at different velocities than the \civ\ variability. Nonetheless, as seen in Fig. \ref{AvT}, the changes in strength in all three marked regions of the \siiv\ trough occur in the same sense as the changes in strength in the two distinct varying regions in \civ.

Another interesting note about 0842+3431 is that we identified it as variable in the short-term, but not in the long-term, in Paper 1. For the long-term comparison in Paper 1, we used the 1990.90 and the 2008.35 observations. However, Figs. \ref{multi} and \ref{AvT} clearly show that the \civ\ BAL varied. The strengh of the BAL is weaker in 2007.04 than in 1990.90, but the BAL becomes stronger again by 2008.35. Therefore, when looking at just the 1990.90 and 2008.35 observations, it appears as if the BAL did not vary at all. This shows how variable profiles can return to a previous state and that variable BALs can be missed in 2-epoch studies.

\subsubsection{0932+5006}
\label{0932}
As in 0146+0142, we note that the absorption trough that appears just redward of the \siiv\ BAL in 0932+5006 in Fig. 1 is actually a high-velocity \civ\ BAL (see Fig. \ref{multi}).

In 0932+5006, there is \civ\ absorption overlapping \siiv\ emission. The variability in these two detectable BAL troughs has a different behaviour than the variability in the other quasars in our sample. When looking at the spectra (Fig. \ref{multi}), the two high-velocity BALs in the 2003.01 spectrum appear to be offset in velocity from the BALs in the other epochs. While this could be indicative of a shift in velocity of the BALs, this apparent offset could also be due to part of the trough weakening while the other part strengthens. This complicates the measurement of $A$ for Fig. \ref{AvT} because any measurement of $A$ in a fixed velocity interval for each of these two high-velocity BALs does not accurately represent how the line changed. Furthermore, the \siiv\ emission line itself could be variable, which complicates any analysis of \civ\ BAL variability in this velocity range.

\subsubsection{0946+3009}
\label{0946}

0946+3009 is the one object in our sample with a redshift too low for our MDM spectra to cover the entire \civ\ region out to the \siiv\ emission line. The spectra only go as blue as $\sim$$-$19000 \kms. The absorption and variability regions that we define for this quasar do not extend all the way to the edge of the MDM spectrum in order to avoid any uncertainties there. The detection of variability in this quasar is secure because we have an additional MDM spectrum of this quasar that matches the MDM spectrum shown in Fig. 1.

\subsubsection{1011+0906}
\label{1011}

As mentioned in Section 2.2, some of the quasars in our sample have low-ionization BALs. We searched for \al\ lines in our sample and used the velocity of the \al\ line to determine the location of \cii. 1011+0906 has an \al\ BAL, but the velocity of the absorption puts \cii\ blueward of the \siiv\ absorption. Therefore, if there is any \cii\ absorption in this object, it does not affect our measurements of the \siiv\ BAL.

The \siiv\ broad emission line (BEL) in this quasar is mostly absorbed by the \civ\ BAL. We fit the \siiv\ BEL using the procedure defined in Paper 1 for cases like this. We take the \civ\ fit, increase the FWHM based on the greater doublet separation in \siiv, and place it at the wavelength where the \siiv\ emission should be. However, there is still some slight emission blueward of the \siiv\ BEL fit. This extra emission may be part of the wing of the \siiv\ BEL, or the underlying power-law continuum fit might be slightly too low. However, the \siiv\ BAL is located at a high enough velocity that any error in the \siiv\ BEL fit should have a negligible effect on our measurements of the BAL and its variability.

\subsubsection{1232+1325}
\label{1232}

1232+1325 has an \al\ BAL which puts \cii\ within the \siiv\ region. The \cii\ BAL is in the velocity range $-$25 500 to $-$19 600 \kms\ (see also Fig. 1), and we omit these velocities from our analysis.

\subsubsection{1303+3048}
\label{1303}

1303+3048 is a BAL quasar that also contained a \civ\ mini-BAL at $\sim$$-$18 500 \kms\ when first observed at Lick (Fig. 1). We only have one Lick observation of this object, but the MDM observations show that the mini-BAL widened and increased in strength to become a BAL. A BAL emerges at the same velocities in \siiv\ as well. The lower-velocity BAL in 1303+3048 is visible in \civ\ in the Lick data, but appears as only weak absorption in \siiv. However, between the Lick and MDM epochs, a \siiv\ BAL emerges and the variability in the \siiv\ absorption extends to lower velocities than the \civ\ variability. The variability at all velocities in this quasar in both \civ\ and \siiv\ occurs in the same sense; the absorption increases in strength.

\subsubsection{1309$-$0536}
\label{1309}

As in 1011+0906, the \siiv\ BEL in 1309$-$0536 is heavily absorbed by a \civ\ BAL. We used the same procedure that we used for 1011+0906 to fit the \siiv\ BEL, and we found what appeared to be significant emission blueward of the \siiv\ BEL fit. In this case, the underlying powerlaw continuum fit did not have the correct slope, so we made a slight adjustment to the continuum fit. Adjusting the powerlaw continuum fit caused on average an increase in $A$ of $\sim$6 per cent throughout most of the \civ\ trough, compared to the measurements in Paper 1. The new \siiv\ BEL fit increased $A$ at the highest velocities in \civ\ by up to a factor of 2. Even with this adjustment, there still appears to be some slight emission blueward of the \siiv\ BEL fit, but, as in 1011+0906, the \siiv\ BAL in 1309$-$0536 is at a high enough velocity that errors in the emission fit should not have much effect on measurements of the BAL. This quasar also did not vary in either \siiv\ or \civ, so any measurement errors for this quasar will not effect any of our results comparing \siiv\ and \civ\ variability properties (e.g., Figs. \ref{dAvdA} or \ref{AAvAA}).

\subsubsection{1331$-$0108}
\label{1331}

As in 1232+1325, 1331$-$0108 has an \al\ BAL at a velocity that places the corresponding \cii\ absorption within the \siiv\ BAL. We therefore omit the velocity range $-$23 800 to $-$15 300 \kms\ in the \siiv\ region from our analysis.

While analyzing the \siiv\ region in 1331$-$0108, we noticed that the pseudo-continuum fit defined in Paper 1 needed to adjusted. Like 1309$-$0536, 1331$-$0108 has very broad BALs, which makes fitting a continuum difficult. The slope of the fit for 1331$-$0108 is now slightly steeper than the fit used in Paper 1, increasing the measured $A$ values for \civ\ on average by just $\sim$7 per cent throughout most of the trough and up to $\sim$30 per cent at the highest velocities, where the absorption is much weaker.

\subsubsection{1423+5000}
\label{1423}

1423+5000 is another quasar where there were two \civ\ BALs that varied, but one BAL increased in strength, while the other weakened. This quasar varied between the Lick and SDSS observations, but we did not detect any variability in the long-term analysis in Paper 1. 0119+0310, 0146+0142, and 1423+5000 are the only quasars in our sample where we see two separate varying regions in \civ\ that did not vary in the same sense.

\subsubsection{1435+5005}
\label{1435}

1435+5005 has \al\ absorption that is either a weak BAL or strong mini-BAL. However, the velocity of the absorption places \cii\ blueward of the \siiv\ absorption.

We also note that the signal-to-noise level in the data for 1435+5005 decreases rapidly at bluer wavelengths. We therefore we do not include the spectral region blueward of $-$12 100 \kms\ in \siiv\ in our analysis in Section 3.1.

\section{Summary of Results}

This is the second paper in a 3-part series to analyze the BAL variabilities in a sample of 24 BAL quasars measured originally by Barlow (1993) at the Lick Observatory in 1988-1992. We supplement those data with spectra from the SDSS archives (for 8 quasars) and our own measurements obtained at the MDM observatory (Table 1). In Paper 1 we discussed the variability properties of \civ\ $\lambda$1549 measured in just two epochs that span a ``short-term" (0.35$-$0.75 yr) and a ``long-term" (3.8$-$7.7 yrs) time interval. Here we build upon that work by including our full multi-epoch data set for these same quasars and making detailed comparisons between the \siiv\ and \civ\ BAL behaviors. Our main results are the following:

\begin{enumerate}
 \renewcommand{\theenumi}{(\arabic{enumi})}
  \item BAL variability usually occurred in only portions of the BAL troughs (Paper 1; Section 3.3.2).
  \item In the long-term interval, 65 per cent of the BAL quasars in our sample showed \civ\ BAL
    variability while only 39 per cent varied in the short-term (Paper 1).  
  \item \civ\ variability occurs more often at higher velocities and in shallower absorption troughs (or
    shallower portions of absorption troughs; Paper 1).  
  \item In rare cases, BAL features appear, disappear, or change to or from narrower mini-BAL
    features (Paper 1; Section \ref{1303}).  
  \item \civ\ BALs in our data are as strong or stronger than \siiv\ BALs at all velocities (in all
    well-measured cases; Fig. \ref{AvA}).  
  \item \siiv\ BALs are more likely to vary than \civ\ BALs. For example, when looking at flow
    speeds $>$$-$20 000 \kms, 47 per cent of the quasars in our sample exhibited \siiv\ variability
    while 31 per cent exhibited \civ\ variability (Section 3.1). The greater variability in \siiv\ is likely
    due to a combination of items (3) and (5) above; weaker lines are more likely to vary, and \siiv\
    tends to be weaker than \civ.  
  \item Variability in \siiv\ can occur without corresponding changes in \civ\ at the same velocities.
    $\sim$50 per cent of the variable \siiv\ regions did not have corresponding \civ\ variability at the
    same velocities. However, in only one poorly-measured case were changes in \civ\ not matched
    by \siiv\ (Sections 3.1 and \ref{0119}).   
  \item At BAL velocities where both \civ\ and \siiv\ varied, the changes always occurred in the same
    sense (Section 3.1).  
  \item We do not find any correlation between the absolute change in strength in \civ\ and in \siiv\
    (Fig. \ref{dAvdA}), but the fractional change in strength tends to be greater in \siiv\ than in \civ\
    (Fig. \ref{AAvAA}).
  \item When additional observing epochs are included (e.g., our full data set; Section 3.2), the
    fraction of \civ\ BALs that varied at any velocity increases from 65 per cent to 83 per cent. This
    increase was caused by variations missed in the 2-epoch comparisons in Paper 1.  
  \item BAL changes at different velocities in the same ion {\it almost} always occurred in the same
    sense (getting weaker or stronger) but not generally by the same amount (Section 3.2). We find
    just 3
    cases that show evidence for one \civ\ BAL weakening while another strengthens within the
    same object (Sections \ref{0119}, \ref{0146}, and \ref{1423}).
  \item The multi-epoch data also show that the BAL changes across 0.04$-$8.2 years in the rest
    frame were not generally monotonic (Section 3.2). Thus, the characteristic time-scale for
    significant line variations, and (perhaps) for structural changes in the outflows, is less than a few
    years.  
  \item With more epochs added, we still do not find clear evidence for acceleration or deceleration
    in the BAL outflows (Section 3.2).  
  
\end{enumerate}

\section{Discussion}

The BAL variability data provide important constraints on the outflow physical properties. However, the information we derive depends critically on what causes the BAL variations. In this section we discuss pros and cons of two competing scenarios, namely, 1) fluctuations in the far-UV continuum flux that cause global changes in the outflow ionization, and 2) outflow clouds moving across our lines-of-sight to the quasar continuum source.

An important part of this discussion is the BAL optical depths, which can be much larger than they appear in the spectrum if the absorbers cover only part of the background light source (\citealt{Hamann98}; \citealt{Hamann08}). Comparisons between the \civ\ $\lambda$1549 and \siiv\ $\lambda$1400 BALs can help because these lines probe slightly different ionizations with potentially very different line optical depths. For example, in a simple situation with solar abundances and an ion ratio equal to the abundance ratio, i.e., \siiv/\civ\ = Si/C, the optical depth in \siiv\ $\lambda$1400 would be $\sim$3.4 times less than \civ\ $\lambda$1549 (\citealt{Hamann97d}; \citealt{Hamann99}; \citealt{Asplund09}). In actual BAL flows, the relative \siiv\ optical depth should be even lower because BAL ionization tends to be high and thus favors \civ. We cannot make specific comparisons without specific knowledge of the absorber ionizations. However, if we reasonably assume that the ionization is at least as high as that needed for a maximum \civ /C ratio (e.g., in a gas that is photoionized by the quasar and optically thin in the Lyman continuum -- fig. A1 in Hamann et al. 2011), then the \siiv\ optical depths should be $>$8 times smaller than \civ.

\subsection{Changing Ionization}

When there is variability in different velocity intervals within the same BAL or within multiple BALs in the same quasar, the changes almost always occur in the same sense (e.g., 0842+3431 and 0903+1734 -- Figs. \ref{multi} and \ref{AvT}). Studies of narrow absorption line (NAL) variability have observed multiple NALs in a given quasar varying in concert (\citealt{Misawa07}; \citealt{Hamann11}). \citet{Hamann11} found coordinated line variations in five NAL systems in a single quasar. They argue that the most likely explanation for this is a global change in ionization. If there are changes in the ionizing flux incident on the entire outflow, then global changes in ionization should occur. While the connection between NALs and BALs is unclear, this argument can be applied to BALs as well. Absorbing regions at different velocities have different radial distances from the central SMBH. They are therefore spatially distinct, even if they are part of the same larger outflow structure. A change in covering fraction due to moving clouds is unlikely in cases such as 0842+3431 and 0903+1734 because it would require coordinated movements among multiple absorbing structures at different outflow velocities and radii.

To further investigate this scenario, for simplicity, we consider a system with homogeneous clouds outflowing from the accretion disk, with no transverse motion across our line-of-sight to the quasar. A change in ionization will cause the optical depths in the lines to change. As mentioned above, the optical depths in \civ\ are higher than in \siiv, so \siiv\ would be more susceptible to changes in ionization. Therefore, it is more likely for \siiv\ to vary than \civ\ in this scenario because \civ\ is more likely to be saturated. This generally matches our results since we find that \siiv\ is more variable than \civ. We find only one case of \civ\ variability unaccompanied by \siiv\ variability, and it is a very tentative result (Section \ref{0119}).

This scenario becomes a little more complicated when considering that typically variability only occurs in portions of BAL troughs, rather than entire BAL troughs varying (\citealt{Gibson08}; Paper 1). As mentioned above, a change in ionization should cause more global changes, rather than changes in small, discrete velocity intervals. It is possible that the variable regions in the troughs have moderate or low optical depths, while the non-variable sections are too saturated to respond to modest changes in the ionization and line optical depths. We have evidence from Paper 1 and Fig. \ref{histA} that weaker lines, or weaker portions of lines, are more likely to vary. However, we also found in Paper 1 that variability is more common at higher velocities, where the absorption tends to be weaker, so it is difficult to say whether it is the higher velocity or the weaker absorption strength that is the root cause of the variability. If it is true that weaker portions of lines, which are least likely to be saturated, are more likely to vary, regardless of outflow velocity, then this would support the changing ionization scenario.

However, if it is true that weaker portions of lines are less saturated and thus more likely to vary, it is unclear why, for example, the weak blue wing of the BAL trough in 0842+3431 does not vary. If changing ionization is causing the variability in this quasar, then the wings of the line must be saturated while the portions of the line adjacent to the wings are not saturated. There are other examples of similar behavior. In 1524+5147, the strongest variability occurs in the deepest segment of the BAL, and there is weak or no variability in the weakest segments of the trough (Figs. 1 and \ref{multi}). Similarly, the variability in 1011+0906 occurred near the core of the line, while the wings did not vary (Fig. 1).

In order for changing ionization to cause variability in just the deepest Êportions of BAL troughs, as in the examples given above, there must be velocity-dependent covering fractions with velocity-dependent optical depths. In this way, even the weak wings can be highly saturated. There is evidence in the literature that both optical depth and covering fractions can have complex velocity-dependent behaviors (\citealt{Barlow97}; \citealt{Hamann97}, \citeyear{Hamann01}; \citealt{Ganguly99}; \citealt{deKool02}; \citealt{Gabel05}, \citealt{Gabel06}; \citealt{Arav08}).

The true optical depths and covering fractions are difficult to measure for BALs. In our data, 1413+1143 provides direct evidence for velocity-dependent optical depths if the line variations are caused by ionization changes. At the core of the \siiv\ trough there are two dips at the \siiv\ doublet separation (Figure 1). The doublet ratio is roughly one-to-one, indicating saturation at these velocities and therefore little or no sensitivity to changes in continuum flux. This part of the trough did not vary. However, this saturated doublet is surrounded by variability at higher and lower velocities in \siiv. In \civ, which should have generally larger optical depths, variability occurs only at higher velocities. This behavior Êis at least suggestive of lower optical depths (non-saturated absorption) at the variable velocities.

One further piece of evidence for the changing ionization scenario comes from the multi-epoch data in Section 3.2, which show that changes in BAL strength are not necessarily monotonic (see also, \citealt{Gibson10}). In fact, in 0842+3431, the absorption trough varied, and then in our last MDM observation it returned to the same strength it had in one of the first Lick observations. In order for a change in covering fraction to have caused the variability in 0842+3431, the cloud movements would have to be repeatable, in addition to being coordinated at different velocities corresponding to different spatial locations. A change in ionization is a more likely explanation because continuum flux variations are not necessarily monotonic either (e.g. \citealt{Barlow93}).

If a change in ionization does indeed cause the variability we detect, then there should be a connection between changes in continuum flux and BAL variability. However, the results from previous studies have been mixed. \citealt{Barlow93} found some evidence for a correlation between continuum variability and BAL variations, at least in certain individual quasars, while other studies have not found a strong correlation (\citealt{Barlow92}; \citealt{Lundgren07}; \citealt{Gibson08}). However, all of these studies look at near-UV flux variations and little is known about the far-UV variability properties of luminous quasars. It is the far-UV flux that is the source of the ionizing radiation. Therefore, these results do not rule out ionization changes as a cause of BAL variability.

\subsection{Changing Covering Fraction}

While most of the evidence presented so far favors ionization changes, previous BAL variability studies, including Paper 1, have favored changing covering fractions over ionization changes (\citealt{Lundgren07}; \citealt{Gibson08}; \citealt{Hamann08}; \citealt{Krongold10}; \citealt{Hall11}). To investigate this possibility, we consider a simple scenario with clouds that have constant ionization and column density, but are moving across our line-of-sight. If \civ\ and \siiv\ have the same covering fraction, then \civ\ should be just as likely to vary as \siiv, which is inconsistent with the results of Section 3.1. Further, the change in strength in the two lines should be the same, which is contradicted by our results in Fig. \ref{dAvdA} (also, \citealt{Gibson10}). Hence, this simple scenario clearly does not match the results of this and previous work.

A more realistic scenario involves clouds that can have different covering fractions in \civ\ and \siiv\ (\citealt{Barlow97}; \citealt{Hamann97}, \citeyear{Hamann01}; \citealt{Ganguly99}; \citealt{Gabel05}, \citealt{Gabel06}; \citealt{Arav08}). \citet{Hamann01} and \citet{Hamann04} discuss simple schematics of inhomogeneous clouds that could lead to different covering fractions in different ions (see fig. 6 in \citealt{Hamann01} and fig. 2 in \citealt{Hamann04}). Stronger transitions in more abundant ions can have a larger optical depth over a larger area in these schematic models. Thus, \siiv\ may trace a different area of the outflowing gas clouds than \civ.

If the \civ\ and \siiv\ lines are saturated, e.g. like the BALs in \citealt{Hamann08}, then the strengths of the lines would be governed by the covering fractions in those lines. In this case, a smaller covering fraction in \siiv\ would be consistent with the results of Fig. \ref{AvA}, which shows that \siiv\ lines are generally weaker than \civ. If \siiv\ is tracing a smaller area of the gas cloud than \civ\ and this cloud is moving across our line-of-sight, then \siiv\ absorption would generally be more variable. Furthermore, if the covering fractions are different for each ion, then the change in covering fraction, as well as the fractional change in strength of the absorption lines, for each ion can also differ. This is consistent with Figs. \ref{dAvdA} and \ref{AAvAA}.

In Sections \ref{0119}, \ref{0146}, and \ref{1423}, we report on three cases that show evidence of one BAL, or one portion of a BAL, strengthening while another weakens within the same quasar. This can be readily explained in a moving cloud scenario, for it is possible for clouds at different velocities to enter/leave our line-of-sight at different times. If one cloud enters our line-of-sight, while another is exiting, we would see one BAL strengthening while another weakens.

\subsection{Conclusions}

The higher variability fractions in \siiv\ versus \civ\ BALs and coordinated variabilities between absorption regions at different velocities in individual quasars supports the scenario of global changes in the ionization of the outflowing gas causing the observed BAL variability. Furthermore, velocity-dependent covering fractions and optical depths could explain why in many cases we see variability in just portions of BAL troughs, rather than entire troughs varying. On the other hand, variability in portions of BAL troughs fits naturally in a scenario where movements of individual clouds, or substructures in the flow, are causing changes in covering fractions in the absorption lines. This scenario is also consistent with the main results of Section 3.1, assuming that \siiv\ has a smaller covering fraction than \civ.

In reality, changes in ionization and covering fractions could both be contributing to BAL variability. In our sample, there are quasars in which we observed no variability; there are quasars that varied in only one or two narrow velocity intervals; and, there are yet other quasars with variations over a wide range in velocities. It is unlikely that one scenario is governing the changes in all of these quasars. Perhaps in quasars where the lines are more saturated, the lines are not susceptible to small changes in ionization, but can easily vary due to covering fraction changes. In other cases, where the lines have lower optical depth, a change in ionization can cause large changes over a wide range in velocities, possibly masking variations due to changes in covering fraction.

There are still some unanswered questions that our results from Paper 1, and the current work, raise. In particular, in Paper 1, we find correlations between incidence of \civ\ variability and both velocity and absorption strength. However, velocity and absorption strength are also correlated. If the trend is really with absorption strength, indicating that weaker lines, which are less likely to be saturated, are more variable, then this favors ionization changes. If the trend is with velocity, then the implications are more ambiguous, but it would be more consistent with the crossing cloud scenario. Clouds with higher outflow velocities are more likely to have greater transverse velocities as well.

There are also documented cases of BALs emerging where there had hitherto been no absorption (\citealt{Hamann08}; \citealt{Krongold10}), and in this work, we report on a quasar (1303+3048; Section \ref{1303}) where a mini-BAL became a BAL. These scenarios speak to the general complexity of BAL variability. Previous studies have hypothesized that different outflow lines may indicate different inclinations of our lines-of-sight to the quasars and that at certain inclinations no absorption is seen (\citealt{Elvis00}; \citealt{Ganguly01}). The connection between BALs, mini-BALs, and NALs is still unclear. If BAL and mini-BAL outflows occur at different inclinations, then perhaps our line-of-sight to 1303+3048 goes through a region of overlap between the mini-BAL and BAL inclinations. One might expect this putative border region between the BAL and mini-BAL parts of the flow to be the most turbulent or unstable, and thus the most prone to showing line variability caused by structural changes/motions in the flow.

This work is only the second paper in a 3-part series on BAL variability. The next paper will include 1) a more thorough exploration of the variability time-scales, with new data added to give extensive coverage across week to month intervals; and 2) a more complete discussion of the implications of variability, e.g., in terms of the size, location, and stability of outflow structures.

\section*{Acknowledgments}

We thank an anonymous referee for helpful comments on the manuscript. Funding for the SDSS and SDSS-II has been provided by the Alfred P. Sloan Foundation, the Participating Institutions, the National Science Foundation, the U.S. Department of Energy, the National Aeronautics and Space Administration, the Japanese Monbukagakusho, the Max Planck Society, and the Higher Education Funding Council for England. The SDSS Web Site is http://www.sdss.org/.

\bibliographystyle{mn2e}

\bibliography{bibliography}

\bsp

\label{lastpage}

\end{document}